\documentclass[twocolumn]{openjournal}
\usepackage{graphicx}
\usepackage{accsupp}
\usepackage{hyperref}
\usepackage{ragged2e}
\usepackage{booktabs}
\hypersetup{breaklinks,colorlinks,citecolor=blue,urlcolor=cyan}
\usepackage{bm}

\newcommand{\orcidauthor}[3]{\author{#2$^{#3}$ \href{http://orcid.org/#1}{\includegraphics[scale=0.04]{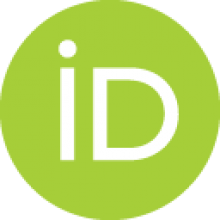}}}}

\usepackage{multirow}
\usepackage{rotating}
\usepackage{array}

\newcommand{\nustar}{NuSTAR }
\newcommand{\relxillcp}{\texttt{relxillCp }}
\newcommand{\kerrbb}{\texttt{kerrbb }}

\begin{document}

\title{Spinning Between Models: Continuum and Reflection Constraints in the Intermediate States of GRS 1716-249 and GRS 1739-278\vspace{-17mm}}

\orcidauthor{0009-0002-4601-6377}{Tiffany Tausch$^1$ }{}
\orcidauthor{0000-0002-2218-2306}{Paul A. Draghis$^{1}$ }{\star}
\orcidauthor{0000-0003-0172-0854}{Erin Kara$^1$ }{}
\orcidauthor{0000-0002-0572-9613}{Abderahmen Zoghbi$^{2,3,4}$}{}
\orcidauthor{0000-0003-2663-1954}{Laura Brenneman$^{5}$}{}
\orcidauthor{0000-0003-4504-2557}{Anna Ogorza{\l}ek$^{2,4,6}$}{}
\orcidauthor{0000-0001-5506-9855}{John A. Tomsick$^{7}$}{}
\orcidauthor{0000-0003-1621-9392}{Mark Reynolds$^{8}$}{}
\orcidauthor{0000-0003-3828-2448}{Javier Garcia$^{6,9}$}{}
\orcidauthor{0000-0002-8908-759X}{Riley Connors$^{10}$}{}


\affil{$^1$ Department of Physics \& MIT Kavli Institute for Astrophysics and Space Research, MIT, Cambridge, MA 02139, USA\\
$^2$ Department of Astronomy, University of Maryland, College Park, MD 20742, USA\\
$^3$ HEASARC, Code 6601, NASA/GSFC, Greenbelt, MD 20771, USA\\
$^4$ CRESST II, NASA Goddard Space Flight Center, Greenbelt, MD 20771, USA\\
$^5$ Center for Astrophysics, Harvard \& Smithsonian, 60 Garden Street, Cambridge, MA 02138, USA\\
$^6$ X-ray Astrophysics Laboratory, NASA Goddard Space Flight Center, Greenbelt, MD 20771, USA\\
$^7$ Space Sciences Laboratory, 7 Gauss Way, University of California, Berkeley, CA 94720-7450, USA\\
$^8$ Department of Astronomy, Ohio State University, 140 West 18th Ave., Columbus, OH 43210, USA\\
$^{9}$ Cahill Center for Astronomy \& Astrophysics, California Institute of Technology, Pasadena, CA 91125, USA\\
$^{10}$ Department of Physics, Villanova University, 800 E. Lancaster Avenue, Villanova, PA 19085, USA}

\thanks{$^\star$E-mail: \href{mailto:pdraghis@mit.edu}{pdraghis@mit.edu}}

\begin{abstract}

Measurements of stellar-mass black hole spin probe accretion under strong gravity and constrain black hole formation and evolution. Intermediate states, in which thermal disk, coronal, and reflection emission all contribute significantly, are particularly complex environments for spin measurements and useful for probing spectral-modeling systematics. We investigate these effects using paired, non-simultaneous Swift/XRT and NuSTAR observations of GRS~1716--249 and GRS~1739--278. We jointly model the spectra with \texttt{kerrbb} and \texttt{relxillCp}, linking or separately varying their spin parameters, and explore the parameter space with a Markov Chain Monte Carlo analysis. Within the adopted model, GRS~1716 strongly favors a high spin, whereas the GRS~1739 data allow high- and low-spin solutions with nearly identical fit statistics. Parameter-stepping experiments show that coordinated changes among parameters allow substantially different physical configurations to produce nearly indistinguishable spectra. For these datasets and within the adopted model, energy-band tests show that the Fe band provides the strongest direct sensitivity to the inferred spin, while the Compton hump helps constrain reflection parameters and the soft X-ray band characterizes the underlying continuum. When both the disk continuum and reflection components are adequately constrained, their associated spin parameters can favor a common solution within the joint model. With weaker constraints, the model permits statistically comparable solutions spanning nearly the full range of spins. Robust joint continuum and reflection spin inference requires broadband coverage, independent binary constraints, physically self-consistent models, and thorough exploration of parameter space.

\end{abstract}

\section{Introduction}
\label{sec:intro}

Stellar mass black holes (BHs) are the natural consequence of the evolution of the most massive stars. By being objects of pure gravity, the properties of space-time in their proximity are entirely determined by the mass and rotation of the BH, the latter generally expressed through the spin parameter $a$. This dimensionless parameter is defined as $a=Jc/GM^2$, where $J$ and $M$ represent the angular momentum and mass of the BH, $c$ represents the speed of light, and $G$ represents the gravitational constant. Values of the parameter range from -1 to 1, with 1 representing a maximally spinning prograde (with respect to the accretion disk) BH, 0 indicating a nonspinning (Schwarzschild) BH, and -1 representing a maximally spinning retrograde BH. Studies of stellar mass BHs offer insight into the fate of massive stars by providing expected test end-products for any computational models aiming to describe supernova processes (see, e.g., \citealt{2025ApJ...987..164B}).  

Transient stellar-mass BHs are most often discovered during periods in which they accrete significant amounts of matter from a stellar companion, forming a hot accretion disk that preferentially emits X-ray radiation. During these transient, short-lived episodes of increased accretion, binary systems containing a compact object -- in this case a BH -- that is feeding on a star become among the brightest sources of X-ray radiation in the sky, hence giving them the name of X-ray binaries (XRBs). \cite{2006ARA&A..44...49R} offer a review of the properties of accreting BH binaries. Studies of XRBs offer a unique avenue to study the evolutionary pathways of double or multiple stellar systems. Furthermore, owing to their brightness and comparatively short variability timescales, stellar-mass BHs provide valuable laboratories for studying accretion physics that can also inform our understanding of accretion onto supermassive BHs, despite the substantial differences in their environments and characteristic scales (see, e.g., \citealt{2005Ap&SS.300..189M, 2007MNRAS.380..301K, 2017MNRAS.466.4121C, 2012MNRAS.419..267P}).

In the literature, most BH spin measurements using X-ray spectroscopy come from one of two main methods: ``continuum fitting" (see, e.g., \citealt{2006ApJ...636L.113S, 2009ApJ...701.1076G}) and ``relativistic reflection" (see, e.g., \citealt{2006ApJ...652.1028B, 2009ApJ...697..900M, 2009MNRAS.395.1257R}). Both methods assume an accretion disk extending to the innermost stable circular orbit (ISCO) of the BH. Furthermore, under certain circumstances, it is possible to apply both methods to the same dataset, as seen in studies such as \cite{2024MNRAS.529.1752D} and \cite{2024A&A...691A.192Z}. This requires significant contribution from both the accretion disk (modeled through the continuum fitting method) and from the reflected radiation (modeled through the relativistic reflection method). At the same time, the spectral overlap of the disk, coronal, and reflection components makes intermediate states less clean environments for either technique than the spectral states in which the methods are traditionally applied. Because intermediate states exhibit both thermal disk emission and coronal activity, modeling only one of these components provides an incomplete physical description of the BH accretion. Although continuum fitting and relativistic reflection have traditionally been applied independently, they describe distinct physical processes that occur simultaneously in intermediate states. We therefore investigate how these complementary modeling techniques can be used in tandem to characterize intermediate-state spectra and constrain black hole spin.

Continuum fitting models the thermal emission from an accretion disk while accounting for relativistic effects in the vicinity of a BH. The method is generally applied to observations obtained during soft, disk-dominated spectral states and requires independent constraints on system properties such as the BH mass, distance, and disk inclination. The observed disk spectrum is commonly described as a color-corrected multi-temperature blackbody, with the spectral hardening factor $f_{\rm col}$ defined as the ratio of the color temperature to the effective temperature. The inferred inner disk radius depends on the measured disk temperature and luminosity, as well as on $f_{\rm col}$ and the system parameters. Under the assumption that the disk extends to the ISCO, this radius can then be used to constrain the BH spin. Therefore, correlations between the spin, $f_{\rm col}$, mass accretion rate, mass, distance, and inclination can contribute substantially to the uncertainty in continuum-fitting measurements \citep{1995ApJ...445..780S, 2000MNRAS.313..193M, 2004MNRAS.347..885G, 2004ApJ...601..428K, 2013MNRAS.431.3510S}. In this work, we model the disk emission using \texttt{kerrbb} \citep{2005ApJS..157..335L}, allowing the spectral hardening factor and the other relevant continuum parameters to vary during the fits.

Relativistic reflection studies model the spectrum of the coronal radiation that is reprocessed by the accretion disk atmosphere. The main features of relativistic reflection are the Fe K complex around 6.4 keV, the Fe K edge, and a broad excess above 20 keV, referred to as the Compton hump. By characterizing the distortion of these spectral features caused by the gravity of the BH, one can infer the proximity of the emission to the BH, and by further equating this distance to the size of the ISCO around a BH, it is possible to directly probe the BH spin. Reflection studies often employ models such as the ones in the \texttt{relxill} family (\citealt{2014ApJ...782...76G, 2014MNRAS.444L.100D}) or the \texttt{reflionx} model (\citealt{2005MNRAS.358..211R}). Reflection fitting is generally applied to observations obtained during hard, corona-dominated spectral states. Unlike continuum fitting, reflection does not require independent constraints on the mass, distance, or disk inclination. 

In an attempt to describe the interplay of models in X-ray spectral characterization of accreting BH XRBs, we analyzed two sets of NuSTAR and Swift/XRT observations taken during the intermediate state of the 2017 outburst of GRS 1716-249 (hereafter GRS 1716) and the 2023 outburst of GRS 1739-278 (hereafter GRS 1739). We chose observations during this spectral state as they provide a balance of flux in the soft band, dominated by the thermal disk emission used in continuum-fitting models, and in the hard band, dominated by the non-thermal coronal emission and radiation reprocessed by the disk atmosphere, used in relativistic reflection models. We chose observations during this spectral state because both the thermal disk emission and the coronal and reflection components contribute appreciably to the observed spectrum. Although this makes intermediate states less ideal for obtaining robust spin measurements with either method individually, it provides a useful testbed for exploring the systematic uncertainties and parameter degeneracies that arise when continuum fitting and relativistic reflection are applied simultaneously. The sources were chosen as they have prior estimates of their distances, and BH masses: GRS 1716: $M_{\rm BH}=6.4\pm3.2\;M{_\odot}$; $d=6.9\pm1.1\;\rm kpc$ (measured using the profile of the H$\alpha$ line in quiescent spectra by \citealt{2023MNRAS.526.5209C}), and GRS 1739: $M_{\rm BH}=6.75\pm2.75\;M{_\odot}$ (estimated based on empirical relation between the luminosity of the transition between spectral states and the Eddington luminosity in \citealt{2018PASJ...70...67W}); $d=7.25\pm1.25;\rm kpc$ (estimated based on the extinction along the line of sight in \citealt{1996A&A...314L..21G}). Furthermore, both sources have estimates of moderate galactic absorption along the line of sight, as estimated by the HEASARC $N_H$ tool\footnote{\url{https://heasarc.gsfc.nasa.gov/cgi-bin/Tools/w3nh/w3nh.pl}}: $\sim2.5\times10^{21}\rm \; cm^{-2}$ for GRS 1716 and $\sim7\times10^{21}\;\rm cm^{-2}$ for GRS 1739.

We use these two datasets as case studies to investigate how spectral properties, data quality, and model flexibility affect joint continuum-fitting and relativistic-reflection spin constraints. This paper is structured as follows. In Section \ref{sec:data} we present the process of extracting spectra from the observations of the two sources. In Section \ref{sec:analysis} we present the spectral analysis. In particular, in Section \ref{subsec:par-relationships} we present a detailed study of correlations between model parameters that can lead the continuum and reflection components to favor different spin values within the same joint fit. In Section \ref{subsec:regions}, we study the impact of different energy bands in the spectra on the ability to constrain BH spin, in an attempt to quantify what spectral properties drive BH spin measurements. In Section \ref{sec:discussion} we discuss the implications of our findings.

\section{Data} \label{sec:data}

In order to obtain broadband coverage of both the thermal disk and hard X-ray spectral regions containing the primary reflection features, we incorporated pairs of observations from NuSTAR (\citealt{2013ApJ...770..103H}) and Swift/XRT (\citealt{2005SSRv..120..165B}). The quality of the available data itself represents an important source of systematic uncertainty in spin measurements: signal-to-noise ratio, spectral resolution, and energy coverage determine how effectively competing spectral solutions can be distinguished. The present observations therefore provide an opportunity to explore how these limitations propagate into the inferred model parameters. Table \ref{tab:observations} presents the times, exposure durations, and ObsIDs of the observations used in our study, for both sources. 

\begin{table}[t]
\centering
\scriptsize
\caption{Intermediate State Observations Information\label{tab:observations}}
\renewcommand{\arraystretch}{1.4}
\begin{tabular}{lcccc}
\hline
Source & Telescope & ObsID & Date & Exp. (ks) \\
\hline
\multirow{2}{*}{GRS 1716}   & NuSTAR    & 90301007002      & 07/20/2017 & 89.3             \\
                                & Swift     & 00088233001      & 07/28/2017 & 3.7              \\
\hline
\multirow{2}{*}{GRS 1739}   & NuSTAR    & 90901323002      & 07/11/2023 & 38.3             \\
                                & Swift     & 00033812090      & 07/08/2023 & 0.7  \\
\hline
\end{tabular}

\begin{minipage}{\linewidth}
\scriptsize
Outburst observation information (ObsID, date, exposure length) for GRS 1716 and GRS 1739 from both Swift and NuSTAR
\end{minipage}
\end{table} 
\begin{figure*}[ht!]
\centering
\includegraphics[width=1.0\textwidth]{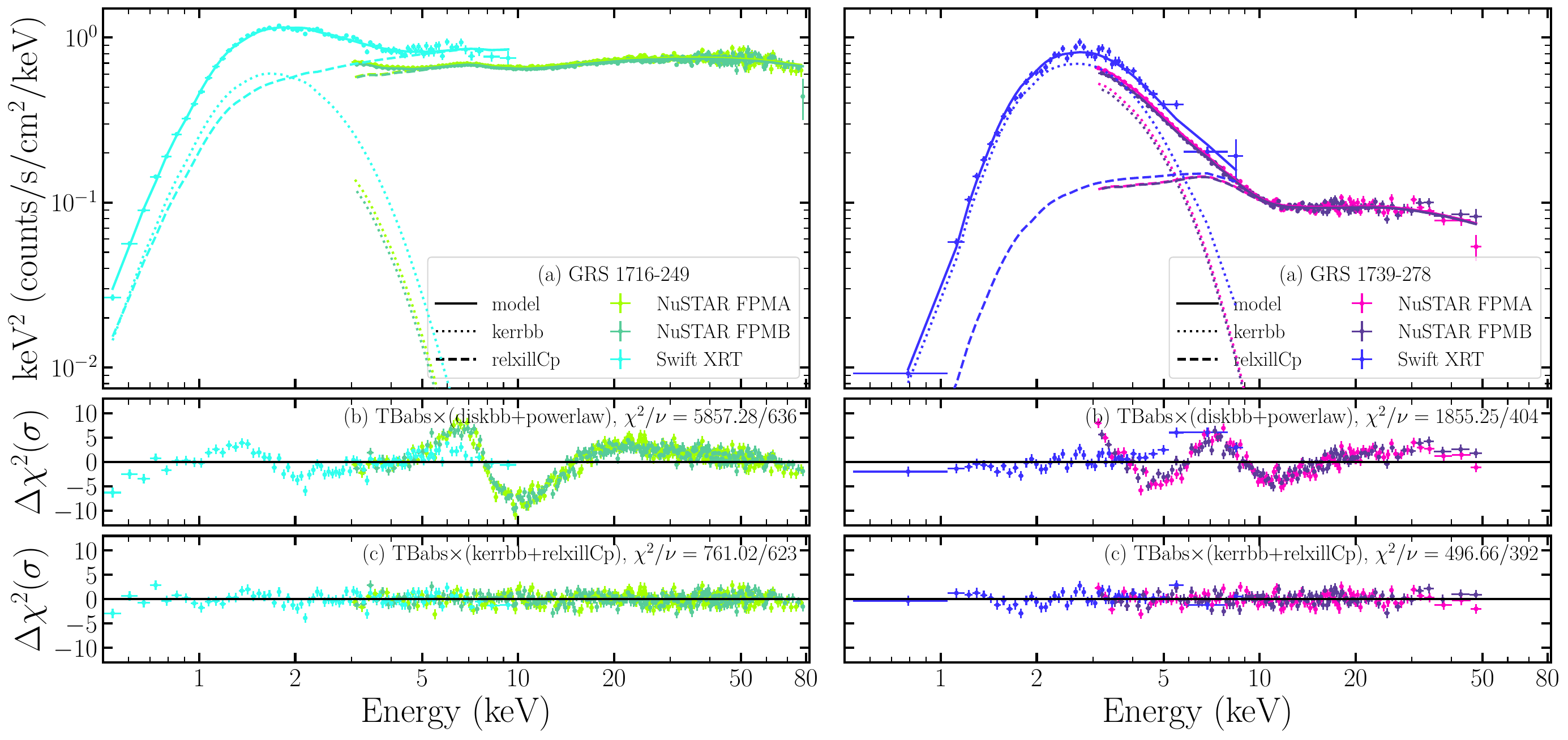}
\caption{Panels (a): Plots of both spectra fit with \texttt{tbabs*(kerrbb+relxillCp)}. When fitting the spectra with \texttt{tbabs*(diskbb+powerlaw)} (Panels b), the residuals in the 5-10 keV band indicate the presence of a broad iron line, motivating the use of relativistic reflection models. Panels (c) show the residuals produced by the best-fit models. The panels on the left show the GRS 1716 data, while the panels on the right show the GRS 1739 data.}
\label{eeplots+residuals}
\end{figure*}

NuSTAR was launched in 2012 and employs two telescopes with 10m focal lengths, allowing them to focus X-rays in the ranges of 3-79 keV. For ObsID 90301007002 (GRS 1716) we extracted source source spectra from circular regions with 120'' radii centered at the position of the source for both NuSTAR focal plane modules (FPMs), and we extracted background regions from annuli with inner radii of 200'' and outer radii of 300'', again centered at the position of the source. For ObsID 90901323002 (GRS 1739), due to stray light contamination, we extracted source and background spectra from circular regions with 120'' radii, placed such that they account for the effects of stray light during the observation. Spectra were extracted using \texttt{nustardas} v2.1.4 and CALDB v.20240520. The light curves of the two NuSTAR observations do not show any strong variability, and therefore we use time-averaged spectra. 

Swift/XRT observations were downloaded from the HEASARC archive\footnote{\url{https://heasarc.gsfc.nasa.gov}} and reprocessed with \texttt{xrtpipeline} to ensure usage of the latest Swift calibration files. Both sources were observed in windowed timing mode. Spectra were extracted from a 90$\times$40 pixel rectangular region centered on the known source position and background spectra were extracted from neighboring annular regions. Due to the source brightness, the background was negligible in both cases. Exposure maps and ancillary response files were generated for each observation using the \texttt{xrtexpomap} and \texttt{xrtmkarf} tasks respectively. Windowed timing mode response files from the latest calibration release were used. 

The net count rate measured from GRS 1739 is 26.33 $\pm$ 0.20 ct/s. The net count rate for GRS 1716 was $\sim$96 ct/s and inspection of the resulting spectrum revealed spectral distortions (e.g., excess structure near the Si K edge) consistent with the presence of mild pile-up. To mitigate pile-up, spectra were subsequently extracted using an annular region excluding the inner 14'' (6 pixels). Although this excision substantially mitigates pile-up, residual pile-up or sensitivity to the chosen excision radius can alter the recovered soft-band spectral shape. We therefore regard the pile-up correction as an additional source of systematic uncertainty in the continuum parameters.

All spectra were fit using \texttt{Xspec} v12.14.1, using the abundances described by \cite{2000ApJ...542..914W} and photoionization cross-sections of \cite{1996ApJ...465..487V}. The NuSTAR spectra of GRS 1716 were fit in the entire 3--79~keV energy band, but the spectra of GRS 1739 were truncated at 50~keV as they became background-dominated at higher energies. The Swift spectra were fit in the 0.5-10~keV band. All spectra were rebinned using the \cite{2016A&A...587A.151K} ``optimal" binning scheme through the \texttt{ftgrouppha} ftool. For plotting purposes, the figures throughout the paper show a stronger binning.

Figure \ref{eeplots+residuals} shows the spectra for GRS 1716 (left) and GRS 1739 (right). The top panels show the data from the different instruments in different colors, and the solid, dotted, and dashed lines show the complete best-fit models, and the contributions of individual model components, respectively, with the different colors corresponding to the associated spectra. The central panels in Figure \ref{eeplots+residuals} show residuals produced when fitting the data with a model that does not account for relativistic reflection, to indicate the presence of reflection features. The lower panels show the residuals in terms of $\sigma$ produced by our best-fit models. 

We jointly fit the spectra of each source, despite the observations being separated by three days for GRS~1739 and eight days for GRS~1716. We obtained MAXI light curves of the two sources and calculated the evolution of their hardness ratios. For GRS 1739, the two observations, separated by three days, occurred at similar hardness and flux. For GRS 1716, where the observations were separated by eight days, the hardness slightly decreased between the NuSTAR and Swift observations, due to an increase in flux in the 2-4 keV band while the 4-10 keV band remained nearly constant. This suggests that the broad hard-band spectral shape did not evolve dramatically between the observations, but does not rule out changes in the disk or reflection spectrum at lower energies. We therefore proceed with a joint analysis of the Swift/XRT and NuSTAR spectra, while treating their non-simultaneity and the observed spectral evolution as an additional source of systematic uncertainty.

\section{Analysis and Observations} \label{sec:analysis}
\subsection{Model configuration}

Each model used to fit the outburst spectra has three components. The first component, \texttt{Tbabs} (\citealt{2000ApJ...542..914W}), accounts for X-ray absorption along the line of sight and has only one parameter, the equivalent hydrogen column density, $N_{\rm H}$, which was allowed to vary freely. The second component, \texttt{kerrbb} (\citealt{2005ApJS..157..335L}), accounts for multi-temperature blackbody emission while incorporating general relativistic effects from the BH. This model includes parameters for the BH mass and distance, which were initialized at the literature values listed in Section \ref{sec:intro} and allowed to vary within hard bounds defined by their reported one-sigma uncertainties. The mass accretion rate and hardening factor were allowed to vary freely, and the normalization of the component was fixed to 1, as recommended when prior information regarding the system properties is used. 

The third component, \texttt{relxillCp}, models the coronal emission and the reflection spectrum from radiation interacting with the accretion disk. We adopted the \texttt{relxillCp} flavor from the \texttt{relxill} (\citealt{2014ApJ...782...76G, 2014MNRAS.444L.100D}) v2.3 family of models as it allows for a variable density of the accretion disk, $\log(n)$. We fixed the inner radius of the accretion disk to the ISCO, the outer radius to $1000\;\rm r_g$, and the redshift $z=0$. All other parameters were allowed to vary freely. However, for the emissivity indices, we set constraints of $q_1\geq3$ and $q_2\leq3$, to match the computational predictions of \citealt{2012MNRAS.424.1284W}, while maintaining the flexibility of the model to converge on a Newtonian solution of $r^{-3}$. Additionally, we constrained the reflection fraction in the model to positive values, to ensure that the model simultaneously includes coronal emission and reflected emission. In \texttt{Xspec} parlance, the complete model is $\texttt{tbabs}\times(\texttt{kerrbb}+\texttt{relxillCp})$. We adopted this additive formulation as a diagnostic model for exploring the interplay between the thermal and nonthermal components, but note that it does not enforce self-consistent transfer of disk seed photons into the Comptonized continuum or subsequent coronal scattering of reflected photons.

To account for cross-calibration uncertainties between detectors and possible source variability between pointings, we allowed the normalizations of the model components to vary between spectra. Because the observations are non-simultaneous, these normalization offsets cannot be uniquely attributed to instrumental calibration and may also absorb intrinsic source variability. For \texttt{kerrbb}, we fixed the normalization of the spectra from the NuSTAR FPMA module to 1, and allowed the normalizations of the models for the spectra from NuSTAR FPMB and Swift/XRT to vary, producing values consistent with known and expected offsets between the detectors. For the \texttt{relxillCp} component, the normalization was allowed to vary freely for each spectrum. Lastly, for the Swift spectra of GRS 1716, to account for any possible pile-up that was not mitigated through the spectral extraction process and for any possible spectral changes between the two observations, we allowed the power-law index $\Gamma$ to vary between the NuSTAR spectra and the Swift spectrum.

Lastly, the BH spin $a$ and the inclination of the inner disk $\theta$ are free parameters in both the \texttt{kerrbb} and the \texttt{relxillCp} components. The inclination is required as a prior in continuum fitting measurements, often being fixed to the binary inclination determined from optical studies of the companion star. However, growing evidence suggests that the inclination of the inner disk can differ from the orbital inclination (see, e.g., \citealt{2025ApJ...995L..12D} on Cygnus X-1). Therefore, we allowed the inclination to vary freely, and linked it between the two components. 

\subsection{Spectral fitting}

We take two approaches to fitting the source spectra. In the first approach, we link the spin parameters between the two model components, and in the second, we unlink the parameters, allowing them to vary separately to test for disagreement. In searching for the best linked spin fit for GRS 1739, we found two distinct solutions: a high-spin and a low-spin fit, both with similar ${\chi}^2$. In the unlinked-spin fit, the two spin parameters diverge: the spin parameter associated with \texttt{kerrbb} converges toward lower values, while that associated with \texttt{relxillCp} converges toward higher values within the same joint fit (specific values of spin and ${\chi}^2$ for each fit are listed in Table \ref{spin_fits}). All three solutions exhibit different values for the spin parameter with comparably low ${\chi}^2$. We discuss the different solutions in the following sections. 

We attempted to find a similar set of solutions for GRS 1716. In the linked spin fit, GRS 1716 appears to favor very high spin values, with any low-spin solutions being associated with a significant increase in ${\chi}^2$, suggesting that they represent local minima in the parameter space. In addition to the high-spin solution ($a\sim0.99$), we find a second local minimum solution with slightly lower spin ($a\sim0.8$) but with $\Delta \chi^2=40$ worse. When the spin is unlinked between the \texttt{kerrbb} and \texttt{relxillCp} solution, the fit converges to a solution in which the spins of both components take the value obtained in the previous high-spin measurement. Alternatively, we identify another local minimum solution with high \texttt{relxillCp} spin and lower \texttt{kerrbb} spin with $\Delta \chi^2=7$ worse than the linked-spin solution. To understand how the models and data drive spin measurements, we chose to continue our study on these different results, despite their worse statistic. 

While reduced ${\chi}^2$ is a widely accepted measure for determining the quality of a spectral fit, there are often nuances that arise in the spectral fitting process, and a fit with low ${\chi}^2$ can correspond to misleading combinations of parameters. It is often possible to discover local minima through incrementally altering parameters, producing unique fits with similarly low ${\chi}^2$. In the case of GRS 1739, altering parameter values or the choice of whether to link parameters between model components produces three spectral fits with similar ${\chi}^2$ and entirely different values for the spin. The lowest ${\chi}^2$ fit has the spin parameter unlinked between \texttt{kerrbb} and \texttt{relxillCp}. However, nearby local $\chi^2$ minima with linked high or linked low spin values are of comparable quality, with statistically insignificant deviations in ${\chi}^2$. Unlike GRS 1739, spectral fits for GRS 1716 behave in a more conclusive manner: the linked high spin solution yields a significantly better fit than the other two scenarios based on ${\chi}^2$, corresponding to a more strongly preferred statistical minimum within the adopted model. Here, we test how the spectral differences between GRS 1739 and GRS 1716 and the properties of our chosen models facilitate the presence of multiple best-fit solutions in the former or a single preferred solution in the latter. 

\subsection{MCMC analysis}

In order to further explore the parameter space of the models, we run a Markov Chain Monte Carlo (MCMC) analysis on the three fits we obtain for each source: the linked high and low spin fits, and the fit where the spin parameter is free between the two components. MCMC chains were run through the Goodman-Weare method (\citealt{2010CAMCS...5...65G}) with 2.5 million steps and 100 walkers per chain. In Table \ref{spin_fits}, we report the spin and ${\chi}^2$ values corresponding to the mode of the posterior distribution from each chain, with the uncertainty in the spin computed by the $1\sigma$ credible interval on either side of this mode. For reference, we include the $\chi^2$ values obtained from the fit. In order to ensure the independence of each step within the chains, we use the methods described by Goodman-Weare (\citealt{2010CAMCS...5...65G}) to compute the maximum autocorrelation time for each fit and use this to appropriately thin the chains prior to computing parameter degeneracies.

\begin{table}[htbp]
\centering
\small
\caption{GRS 1716 and GRS 1739 Spin Fits\label{spin_fits}}
\renewcommand{\arraystretch}{1.4}
\begin{tabular}{lcccc}
\hline
Source & Model & $a$ & ${\chi}^2_{\rm MCMC}$ & ${\chi}^2/$dof \\
\hline
\multirow{4}{*}{GRS 1739} & Linked (High)    & $0.97\pm{0.01}$ & 510 & 497/392                  \\ \cline{2-5}
                          & Linked (Low)   & $0.05^{+0.08}_{-0.24}$ & 510 & 497/392       \\ \cline{2-5}
                          & \texttt{relxillCp} & $\geq0.99$ & \multirow{2}{*}{509} & \multirow{2}{*}{495/391}\\
                          & \texttt{kerrbb}    & $0^\dagger$ &                      \\
\hline                          
\multirow{4}{*}{GRS 1716} & Linked    & $\geq0.99$ & 779 & 761/623                  \\ \cline{2-5}
                          & Linked    & $0.80^{+0.05}_{-0.09}$ & 823 & $811/623^*$                  \\ \cline{2-5}
                          & \texttt{relxillCp} & $\geq0.99$ & \multirow{2}{*}{787} & \multirow{2}{*}{$768/622^*$}\\
                          & \texttt{kerrbb}    & $0.73^{+0.08}_{-0.20}$ &                      \\
\hline
\end{tabular}

\begin{minipage}{\linewidth}
\scriptsize
Linked high, low, and unlinked spin fit for both sources. The ${\chi}^2_{\rm MCMC}$ column contains the mode of the posterior distributions from MCMC simulations and the ${\chi}^2/$dof column indicates the best-fit values and the number of degrees of freedom.

$^\dagger$ Despite being allowed to vary freely, the spin in the \texttt{kerrbb} component in the fit with unlinked spins for the GRS 1739 data does not deviate significantly from zero. See Figure \ref{fig:corner_1739}.

$^*$ These solutions represent local statistical minima, and were only include in our work to enable a study of model properties.
\end{minipage}
\end{table}

\subsection{Parameter Relationships} \label{subsec:par-relationships}

\begin{figure*}[ht]
\centering
\includegraphics[width=\textwidth]{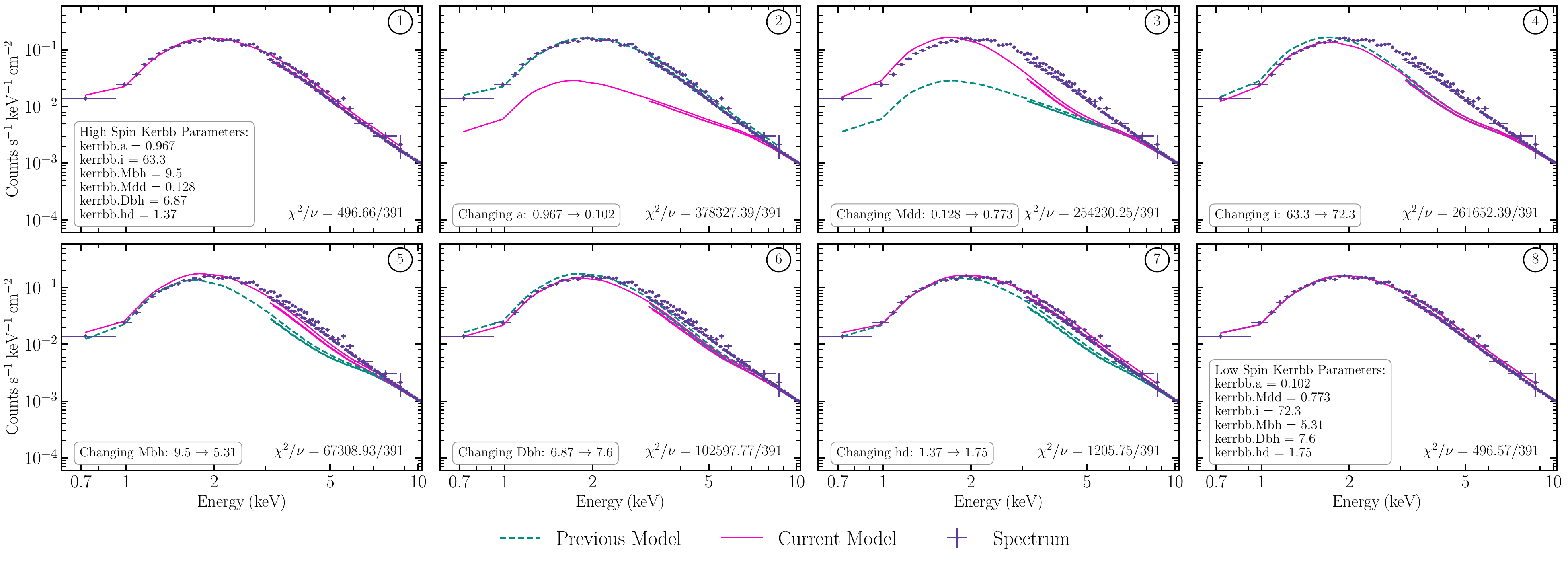}
\caption{Progression of \texttt{kerrbb} parameters from the GRS 1739 linked high spin fit to the linked low spin fit. Graphical insets denote which parameter has changed in the fit progression, and labels in the upper right denote the figure progression. The improvement in fit in the final panel is accomplished by slight changes to the \texttt{relxillCp} component and normalizations in order to better fit higher energy ranges. The initial worsening of fit and progressive improvement illustrates the presence of parameter degeneracies. For instance, the initial decrease in spin and subsequent increase in mass accretion rate return the model closer to its original configuration.}
\label{high_to_low}
\end{figure*}

The existence of spectral fits with substantially different parameter values yet comparable $\chi^2$ suggests that complex correlations among multiple parameters can allow distinct physical solutions to provide similarly good descriptions of the data. Although some of these correlations can be visualized as two-parameter degeneracies, the full behavior of the model is multidimensional, with different parameters affecting different portions of the spectrum.

\subsubsection{From one solution to another}\label{subsubsec:stepping}

To explore possible degeneracies in the \texttt{kerrbb} model component, we examined the impact of changing free \texttt{kerrbb} parameters when shifting from GRS 1739's high-spin fit to the low-spin fit, as presented in Figure \ref{high_to_low}. Without refitting, we started from the high-spin solution and changed the values of the parameters to those obtained in the low-spin solution and evaluated the change in the spectral model and in the fit statistic. Initially, changing the spin parameter substantially changes the fit and increases ${\chi}^2$, but as each parameter alteration takes effect, the model gradually converges to a solution that accurately fits the spectra. The ${\chi}^2$ in the seventh (second from the bottom right) panel highlights the quality of fit after implementing all six parameter changes, and the final frame presents the ${\chi}^2$ after incorporating minor adjustments to the \texttt{tbabs} and \texttt{relxillCp} components to achieve the best-fit low spin solution. For reference, we provide a video that shows the evolution of the model as the parameters transition between the two solutions, which can be found in the online version of the article and at the following \href{https://www.pdraghis.com/research/parameter-stepping}{link}\footnote{\url{https://www.pdraghis.com/research/parameter-stepping}}. Additionally, Table \ref{stepping-endvalues} in Appendix \ref{sec:par_values} shows all the values of the parameters in these two solutions. We note that the high-spin solution for GRS~1739 drives the BH mass to $M_{\rm BH}=9.5\,M_\odot$, the upper boundary of the adopted literature-based range, indicating that this solution is sensitive to the assumed external constraint on the BH mass.

This experiment demonstrates that for the parameters in \texttt{kerrbb}, especially when simultaneously considering the reflection spectrum, it is not sufficient to alter a combination of two parameters to obtain equivalent solutions. It is not sufficient to change, for example, just the hardening factor to compensate for a change in spin. While these degeneracies are known, they are complex and require all parameters to be adjusted simultaneously to produce similar fit statistics. 

Investigating the corner plots in Figures \ref{fig:corner_1716} and \ref{fig:corner_1739} in Appendix \ref{sec:corner_plots} provides initial insights into how the sets of parameters change between the different solutions. In particular, for GRS 1739 (Figure \ref{fig:corner_1739}), all solutions produce similar distributions of $\chi^2$, suggesting that the solutions are comparable in terms of significance. However, the combinations of parameters, and most interestingly the BH spin, take a range of values. For example, the high-spin solution (orange) measures a high power-law index $\Gamma\sim2.8$, whereas the low-spin solution (blue) prefers a lower power-law index $\Gamma\sim2.1$. This inability to definitively constrain the power-law index is likely associated with the source being in a softer, disk dominated spectral state at the time of the observation. Furthermore, the high- and low-spin solutions measure different ionizations, inner emissivity profiles, disk densities, and viewing inclinations. The high-spin solution finds a lower ionization, higher inner emissivity $q_1$, higher disk density ($\log(n)\sim19$ vs. $\log(n)\sim15$ in the low-spin solution), and a lower inclination. It is likely that this combination of parameters allows the reflection model to similarly fit the shape of the broad Fe line with two vastly different spins, with the correlation between the ionization and disk density being discussed in works such as \cite{2023ApJ...951..145L}. Looking at the \texttt{kerrbb} parameters, the two solutions measure disagreeing BH masses, accretion rates, and hardening factors. In particular, the high-spin solution requires a low hardening factor $f\sim1.4$, whereas the low-spin solution prefers higher values of $f$, extending between 1.5-2.5. We note the well-known, particularly strong correlations between the hardening factor and the BH mass, and the hardening factor and BH spin. Other interesting correlations emerge from the intermediate state observation of GRS 1716 (Figure \ref{fig:corner_1716}), between the inclination and inner emissivity $q_1$, inclination and ionization, inclination and reflection fraction, and power-law index $\Gamma$ and ionization. These correlations have previously been reported on by \cite{2025ApJ...989..227D}.

\subsubsection{Alternative coronal geometries}
A notable point of tension emerges in the GRS 1739 low-spin fit: the interplay between the inner disk radius ($R_{\rm in}$) and the breaking radius  ($R_{\rm br}$). By lowering the spin parameter to $0.05^{+0.08}_{-0.24}$, the inner disk radius shifts outward to values slightly below 6 $r_{\rm g}$. Since the breaking radius occurs at 6.1 $r_{\rm g}$ in this low-spin fit, the inner emissivity index ($q_1$) is left with a very small region of the accretion disk to describe. As a result,  $q_1$ is driven to the very upper end of its range ($\approx{9.99}$) in the fit, but the MCMC analysis reveals that it is poorly constrained. Such high values of $q_1$ are usually associated with high-spin black holes and their ability to cause stronger light bending and reflection. This high value of $q_1$ therefore indicates a potentially nonphysical model description. 

In practice, it is likely that in systems with low spins, the broken power-law emissivity profile would be challenging to characterize, even in observations with good signal-to-noise ratios (SNRs). For an observation such as the one presented here, a model that assumes a lamp-post coronal geometry or a simple power-law emissivity profile may be more physically adequate and statistically motivated. We tested this by fitting the GRS 1739 spectra with a model that uses a lamp-post geometry (\texttt{relxilllp}), with a model that fixes $q_1=q_2=3$ and the breaking radius to an arbitrary value in \texttt{relxillCp}, and with a model that combines the lamp-post geometry with the flexibility of a variable disk density \texttt{relxilllpCp}. Similarly to the fits with a coronal free emissivity profile, we attempted to identify high- and low-spin solutions. For the power-law emissivity profile with $q_1=q_2=3$, the fit produces $\chi^2/\nu=504/395$ with a spin of $a\sim0.97$ and $\chi^2/\nu=511/395$ for a spin of $a\sim0$. For the fits with a lamp-post geometry, the fits produce $\chi^2/\nu=501/395$ with a spin of $a\sim0.97$ and $\chi^2/\nu=500/395$ for a spin of $a\sim0$. Lastly, for the \texttt{relxilllpCp} model, the fit produces $\chi^2/\nu=497/394$ with a spin of $a\sim0.97$ and $\chi^2/\nu=496/394$ for a spin of $a\sim0$. The lamp-post geometry, regardless of whether through the \texttt{relxilllp} or \texttt{relxilllpCp} flavor, produces results consistent with the \texttt{relxillCp} fits with free emissivity, with two solutions with nearly identical statistics and inconsistent spin. On the other hand, the model in which we set the emissivity profile to a power law in \texttt{relxillCp} favors a high-spin solution by $\Delta\chi^2=7$ for no difference in the number of degrees of freedom. This experiment confirms that the two solutions identified by our analysis of these spectra are not an artifact of the adopted coronal emissivity profile. 

\subsubsection{Weighted least-squares regression}
Corner plots provide a visual technique to understand the relationship between parameter pairs, with contour lines containing equal probabilities of parameter combinations. However, it is challenging to visualize multi-dimensional parameter correlations. The weighted least-squares regression is a useful tool for understanding these higher-dimensional correlations, and it can be used to evaluate the dependence of one variable on a combination of multiple independent variables. This method is detailed in Appendix \ref{sec:regression_results}, together with some examples. Table \ref{regression-results} in Appendix \ref{sec:regression_results} presents the results of the analysis of the correlation between the spin and sets of 1-4 parameters of interest for all three fits of both sources, together with the associated $R^2$ values, which quantify the strength of the correlation. For the fits with the spin unlinked between the \texttt{kerrbb} and \texttt{relxillCp} components, we present the correlations of the spin parameters in both components. 

We employed weighted least squared regression with combinations of up to four parameters that impact spin, finding the following. Across the different spectral fits, the spin parameter generally shows its strongest correlations with parameters associated with either \texttt{kerrbb} or \texttt{relxillCp}, rather than with combinations spanning both components. The only exception is the low-spin fit of GRS~1716, where some four-parameter combinations containing parameters from both components yield high $R^2$ values. In the lower-spin solutions, the spin shows stronger correlations with the \texttt{kerrbb} parameters. Conversely, in the unlinked-spin fits, the spin parameter associated with \texttt{relxillCp} remains near the high-spin end of the allowed range. These trends suggest that lower-spin solutions are more strongly associated with variations in the continuum parameters within the adopted joint model.

Achieving a lower spin value requires the \texttt{kerrbb} model to ``drive''  the spectral fit, as the \texttt{relxillCp} component tends to statistically favor high spins in the joint fits with unlinked spins. Here and throughout, references to a fit being ``driven'' by \texttt{relxillCp} refer to the full model component, which includes both the direct Comptonized continuum and reflected emission, rather than to the reflection spectrum alone. However, \texttt{relxillCp} is not the only provider of high spin values, as evidenced in the GRS 1739 high-spin fit, where its spin parameter shows its strongest correlations with \texttt{kerrbb} parameters. Overall, we find that the \texttt{kerrbb} component exhibits more flexibility to span a wide range of spin values, spin parameter associated with \texttt{relxillCp} remains above 0.9 in the unlinked-spin fits of both sources.

In the case of GRS~1716, the high spin value associated with \texttt{relxillCp} strongly influences the joint solution, while the \texttt{kerrbb}-associated spin parameter reaches $a=0.73^{+0.08}_{-0.20}$ in the local-minimum unlinked-spin fit. Lowering the overall spin therefore incurs a significant statistical penalty. In contrast, the \texttt{relxillCp} component in GRS 1739 does not influence our spectral fits in this dominating manner, allowing for viable solutions with a range of spin values. Nevertheless, we still see \texttt{relxillCp} exhibit high spin values when the parameter is unlinked across the components. Interestingly, although the $\chi^2$ value in the linked low-spin fit is still comparable to that of the other two GRS 1739 fits, the influence of \texttt{relxillCp} manifests differently: the spin ($a = 0.05^{+0.08}_{-0.24}$) and many other parameters become notably unconstrained. This suggests that internal tensions between model components introduce parameter uncertainties despite the overall fit quality remaining high.

Three of the four spin values in the unlinked cases exhibit abnormally low $R^2$ values, even when running regression with combinations of four parameters. This is likely due to the nature of the calculation for the coefficient of determination. In all three cases, the spin is much more constrained in the MCMCs distributions than in other fits, leaving little intrinsic variation for $R^2$ to capture. When both the weighted squared errors (WSE) and total squared errors (TSE) approach zero, their ratio becomes an indeterminate form with a limit that approaches 1. This can result in a misleadingly low value for $R^2$ despite a best-fit line that might closely approximate the data.

In the fits where the spin shows its strongest correlations with \texttt{kerrbb} parameters, all of the free continuum parameters (the mass $M_{\rm BH}$, mass accretion rate $\dot{M}$, distance $D_{\rm BH}$, hardening factor $f$, and inclination $\theta$) prove to be contributors. Compared to \texttt{relxillCp}, \texttt{kerrbb} has fewer free parameters in our spectral fits, so this observed degeneracy exists across the entire model. As seen in Figure \ref{high_to_low}, lowering the spin from a high to low value initially flattens out the model, but most noticeably by increasing the $\dot{M}$ and $f$, and decreasing $M_{\rm BH}$, the model can largely be brought back to fitting the spectra appropriately. 

In the fits where the spin shows its strongest correlations with \texttt{relxillCp} parameters, the five most frequent contributors to the highest-$R^2$ combinations are the inner emissivity index $q_1$, photon index $\Gamma$, ionization $\log(\xi)$, iron abundance $A_{\rm Fe}$, and reflection fraction. The photon index $\Gamma$ characterizes the shape of the Comptonized continuum, while the ionization parameter modifies the reflected spectral shape, including the Compton hump, through changes in the ionization balance, opacity, and thermal structure of the disk atmosphere. The significance of $A_{\rm Fe}$ likely arises from its impact on the shape of the broad iron line, but also for its ability to break model degeneracies with other parameters when coverage of the Compton hump is available, which in turn impacts the measured spin. The relationship between $q_1$ with the spin parameter is likely to do with the link between the radius of the inner disk ($R_{\rm in}$) and the spin: as the spin increases, $R_{\rm in}$ decreases, effectively altering the spatial area of the disk that the inner emissivity covers and necessitating changes in its values to accommodate the same flux levels (\citealt{2014MNRAS.439.2307F}). This correlation has been reported on previously by works such as \cite{2023ApJ...954...62D, 2025ApJ...989..227D}. Finally, the reflection fraction measures the ratio of direct to reflected photon flux from the BH, and increases in this parameter scale the model upward in flux.
\clearpage
\subsection{Spectral Region Analysis} \label{subsec:regions}

To better understand why the spin parameter shows stronger dependence on one model over another, we performed three experiments focused on how different spectral regions impact important model parameters: spin, inclination, iron abundance, and ionization. In the first experiment, we remove the energy region from 3-10 keV to understand the constraining ability of the iron line. In the second experiment, we remove any energy bins beyond 10 keV, to test the impact of the Compton hump. And lastly, we remove energies from 0.5-3 keV (by excluding the Swift data) to test the ability of NuSTAR data to constrain parameters on its own. We note that we tested this behavior for both high- and low-spin solutions, but in the analysis of GRS 1716, the low-spin fits repeatedly converged to the high-spin solution. Therefore, for both sources, we only report the results of the analysis on the high-spin solutions. These band-exclusion tests probe the constraining information provided by different spectral regions to the combined model, but they do not isolate the energy ranges associated with the individual additive components, which overlap substantially across the spectrum.

\begin{figure*}[ht]
\centering
\includegraphics[width=0.9\textwidth]{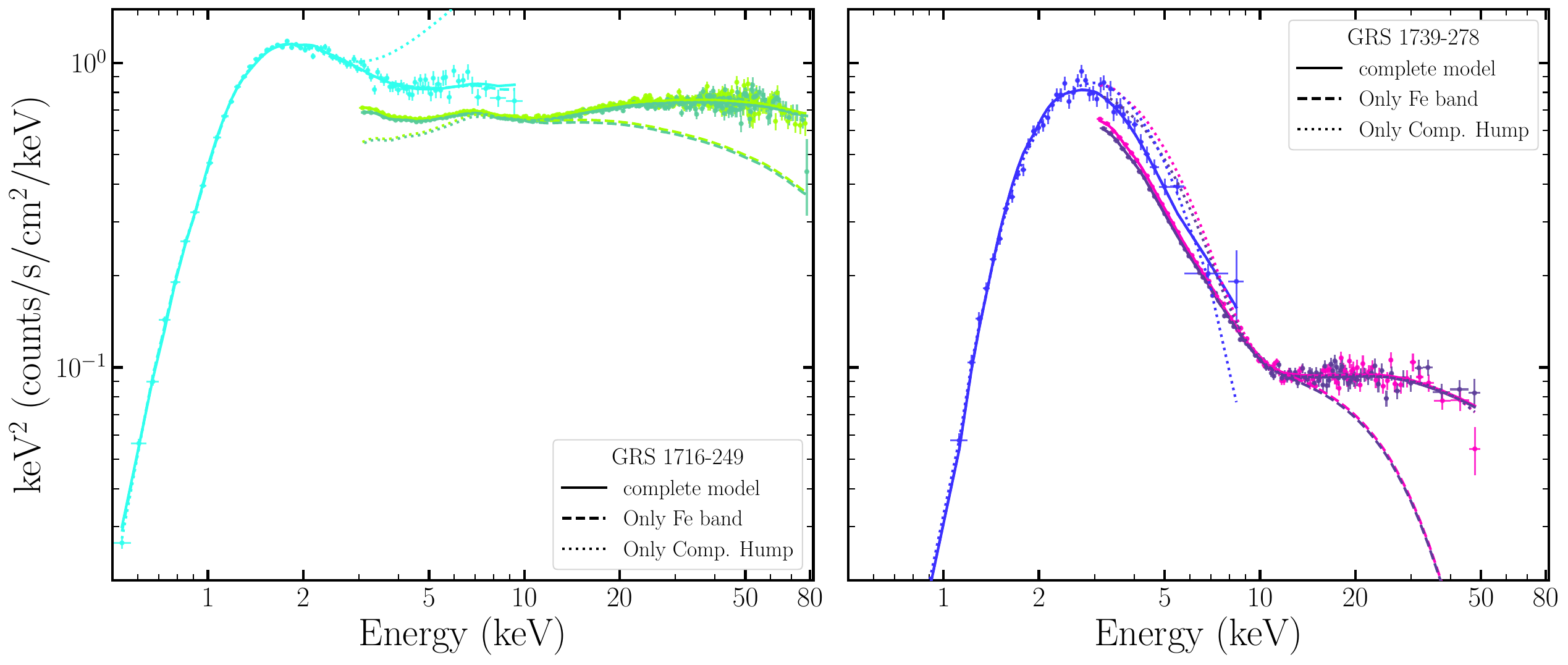}
\caption{Comparisons of models obtained when fitting only certain parts of the spectra and then extrapolating to the entire energy range. The solid lines represent the best-fit models obtained by fitting the entire energy range. The dotted lines show the model produced when only fitting the Compton hump. The dashed lines show the model produced when only fitting the Fe band.}
\label{ignore_and_renotice}
\end{figure*}

Using \texttt{Xspec}’s \texttt{error} command, we first computed the one-sigma confidence intervals of the spin, inclination, iron abundance, and the ionization parameter in our initial spectral fits. After the initial \texttt{error} scans, we removed each spectral region and considered two fitting strategies. In the first, we initialized the fit from the broadband best-fit solution to test how the parameter constraints change when spectral coverage is removed. In the second, we reset the model to generic starting values and refit, testing whether the restricted dataset independently recovers the same region of parameter space. With each fit optimized, we rerun the \texttt{error} command to find out how the exclusion of data influences parameter uncertainties. Lastly, we reincorporate the removed spectral bins to evaluate how the updated models perform on the full data (Figure \ref{ignore_and_renotice}). For both sources, first ignoring an energy band, fitting, and then re-noticing it produces models that diverge significantly from the data. This experiment highlights the importance of obtaining simultaneous broadband coverage of the soft and hard bands.

\begin{figure}
\centering
\includegraphics[width=\columnwidth]{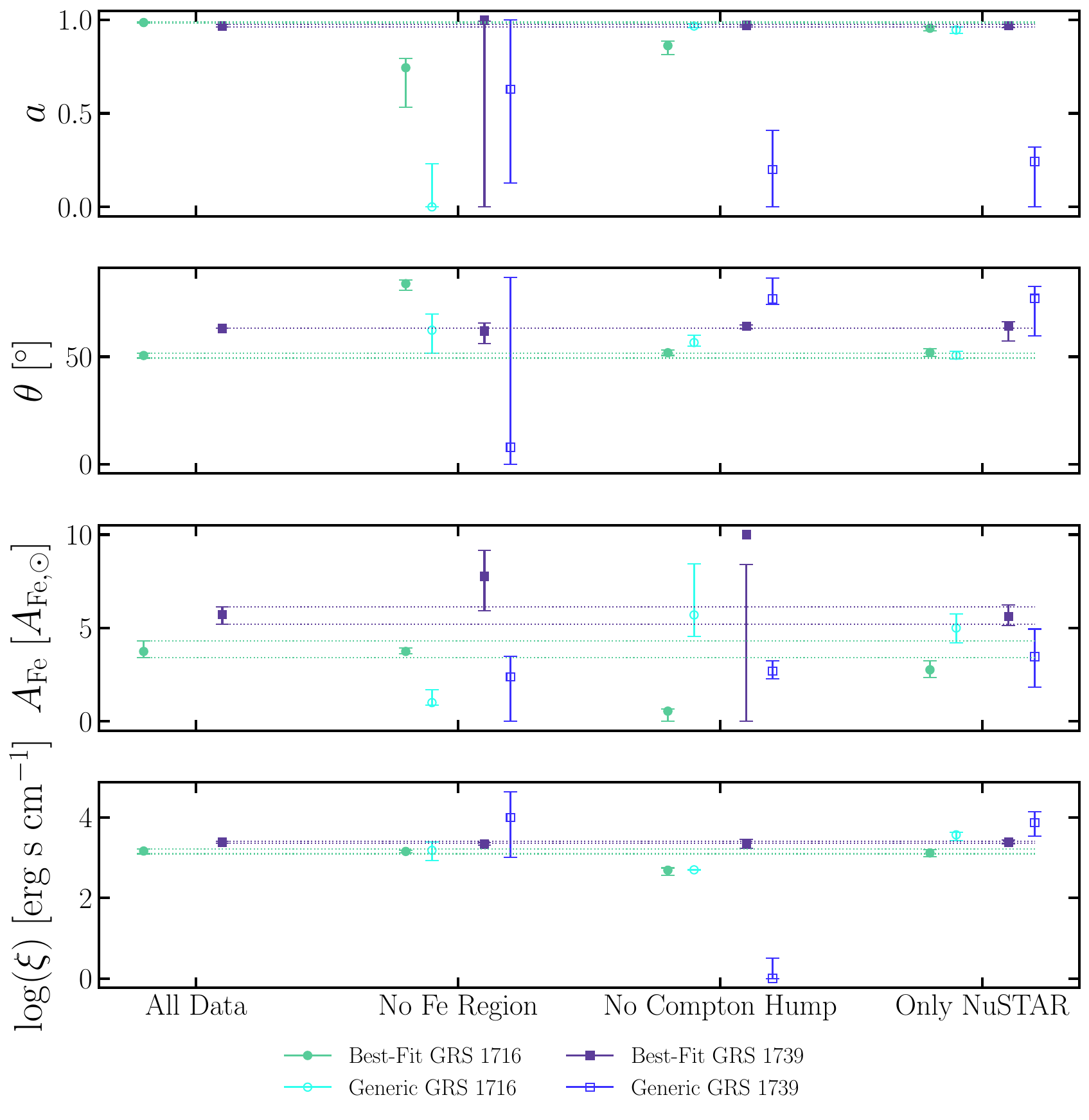}
\caption{Parameter values for spin, inclination, iron abundance, and ionization along with their corresponding one-sigma uncertainties from the \texttt{Xspec} \texttt{error} command. Error bars extend from the original spectral fit containing all data to highlight changes in parameter values and constraints.}
\label{errorbars}
\end{figure}

The results from the \texttt{error} scans on the four key parameters are highlighted in Table \ref{errortable} and Figure \ref{errorbars}. In the case where GRS 1739 starts from a generic fit (empty squares), the spin tends to drift towards lower values while being largely unconstrained, regardless of which energy band is used for the fit, consistent with the continuum parameters playing an important role in determining which region of spin parameter space is recovered. This ultimately impacts the other parameters as well, resulting in higher $\theta$ and lower $A_{\rm Fe}$ values on average. As discussed in the previous section, the overall weakening in constraint might be a result of disagreement between \texttt{relxillCp} and \texttt{kerrbb}. Interestingly, fitting just the NuSTAR spectrum converges to the previously found low-spin solution, highlighting the importance of not only covering the reflection features, but also of the soft part of the spectrum. When starting the fits from the already existing combination of parameters of the high-spin solution (full squares in Figure \ref{errorbars}), all parameters remain around the same values, but the uncertainties increase significantly. In particular, measuring the Fe abundance proves particularly difficult when lacking spectral coverage of the Compton hump.

\begin{table*}[ht]
\centering
\scriptsize
\caption{Parameter Values and Constraints with Spectral Regions Removed \label{errortable}}
\resizebox{\textwidth}{!}{%
\renewcommand{\arraystretch}{1.4}
\begin{tabular}{lclllll}
\hline
Source & Parameter & All Data Range & Fit Start & No Fe Region & No Compton Hump & Only NuSTAR \\
\hline
\multirow{10}{*}{GRS 1716} & \multirow{2}{*}{$a$}                                & \multirow{2}{*}{$0.985^{+0.004}_{-0.002}$}  & Best-Fit  & $0.74^{+0.05}_{-0.21}$  & $0.86^{+0.02}_{-0.05}$   & $0.96^{+0.01}_{-0.02}$   \\
                           &                                                     &                                             & Generic   & $0*$     & $0.967^{+0.012}_{-0.009}$   & $0.96^{+0.02}_{-0.02}$   \\
                           & \multirow{2}{*}{$\theta\;[^\circ]$}                 & \multirow{2}{*}{$51^{+1}_{-1}$} & Best-Fit  & $84^{+2}_{-3}$  & $52^{+1}_{-1}$  & $52^{+2}_{-2}$  \\
                           &                                                     &                                             & Generic   & $62^{+7}_{-11}$ & $57^{+4}_{-2}$  & $51^{+2}_{-2}$  \\
                           & \multirow{2}{*}{$A_{\rm Fe}\;[A_{\rm Fe, \odot}]$}  & \multirow{2}{*}{$3.7^{+0.6}_{-0.3}$}  & Best-Fit  & $3.7^{+0.2}_{-0.1}$   & $0.5^{+0.1}_{-0.5}$   & $2.8^{+0.5}_{-0.4}$   \\
                           &                                                     &                                             & Generic   & $1.0^{+0.7}_{-0.1}$   & $6^{+3}_{-1}$ & $5.0^{+0.8}_{-0.8}$   \\
                           & \multirow{2}{*}{$\log(\xi)\;[\rm erg\;s^{-1}\;cm]$} & \multirow{2}{*}{$3.17^{+0.05}_{-0.07}$}  & Best-Fit  & $3.16^{+0.03}_{-0.04}$   & $2.69^{+0.06}_{-0.12}$   & $3.11^{+0.08}_{-0.09}$   \\
                           &                                                     &                                             & Generic   & $3.2^{+0.2}_{-0.2}$   & $2.699^{+0.002}_{-0.003}$   & $3.57^{+0.06}_{-0.15}$   \\
                           & \multirow{2}{*}{${\chi}^2/\mathrm{dof}$}            & \multirow{2}{*}{$1.22$}                     & Best-Fit  & $1.16$                      & $1.28$                      & $1.17$                      \\
                           &                                                     &                                             & Generic   & $1.21$                      & $1.23$                      & $1.17$                      \\                        
                           \hline
\multirow{10}{*}{GRS 1739} & \multirow{2}{*}{$a$}                                & \multirow{2}{*}{$0.967^{+0.007}_{-0.007}$}  & Best-Fit  & $1*$   & $0.967^{+0.003}_{-0.007}$   & $0.970^{+0.007}_{-0.011}$   \\
                           &                                                     &                                             & Generic   & $0.6^{+0.4}_{-0.5}$   & $0.2^{+0.2}_{-0.2}$   & $0.24^{+0.08}_{-0.24}$   \\
                           & \multirow{2}{*}{$\theta\;[^\circ]$}                 & \multirow{2}{*}{$63^{+4}_{-2}$} & Best-Fit  & $62^{+4}_{-6}$  & $64.2^{+0.5}_{-1.2}$  & $64^{+2}_{-7}$  \\
                           &                                                     &                                             & Generic   & $8^{+79}_{-8}$  & $77^{+10}_{-2}$  & $77^{+6}_{-17}$ \\
                           & \multirow{2}{*}{$A_{\rm Fe}\;[A_{\rm Fe, \odot}]$}  & \multirow{2}{*}{$5.7^{+0.4}_{-0.5}$}  & Best-Fit  & $8^{+1}_{-2}$   & $10*$   & $5.6^{+0.6}_{-0.5}$   \\
                           &                                                     &                                             & Generic   & $2^{+1}_{-2}$   & $2.7^{+0.5}_{-0.4}$   & $3.5^{+1.5}_{-1.6}$   \\
                           & \multirow{2}{*}{$\log(\xi)\;[\rm erg\;s^{-1}\;cm]$} & \multirow{2}{*}{$3.38^{+0.02}_{-0.02}$}  & Best-Fit  & $3.34^{+0.03}_{-0.03}$   & $3.3^{+0.1}_{-0.1}$   & $3.38^{+0.06}_{-0.03}$   \\
                           &                                                     &                                             & Generic   & $4.0^{+0.6}_{-1.0}$   & $0*$    & $3.9^{+0.3}_{-0.3}$   \\
                           & \multirow{2}{*}{${\chi}^2/\mathrm{dof}$}            & \multirow{2}{*}{$1.27$}                     & Best-Fit  & $1.14$                      & $1.58$                      & $1.18$                      \\
                           &                                                     &                                             & Generic   & $1.14$                      & $1.53$                      & $1.17$                 \\
\hline
\end{tabular}
}

\begin{minipage}{\textwidth}
\scriptsize
Results from \texttt{Xspec} \texttt{error} scans on four parameters (spin, inclination, iron abundance, and ionization) for an original fit with all data, as well as for best-fit and generic fits (started with parameter values in the very middle of their ranges and refit) with spectral regions removed.

$^*$ Indicates failed \texttt{error} measurements that are highly unconstrained, with distributions contained close to a boundary value or error bars outside of the initial measurement.
\end{minipage}
\end{table*}

For GRS~1716, removing the Fe region allows the fit to migrate toward a different spin solution, demonstrating that this band provides the strongest direct contribution to the inferred spin. Removing the Compton hump instead weakens constraints on several reflection parameters and on the spin, demonstrating its importance for breaking degeneracies within the broader model. In other words, the Fe band provides the strongest direct sensitivity to the spin, while the Compton hump helps constrain the broader reflection model and break parameter degeneracies, improving the robustness of the inferred spin.

The iron line appears to be the most significant region for constraining the spin, with parameter variation increasing by an average factor of 50 when removed compared to including all data. This region also proves to be important for constraining both $\theta$ and $A_{\rm Fe}$, while values and constraints on $\log(\xi)$ remain largely unaffected. The dependence of the iron abundance $A_{\rm Fe}$ on the iron region is not surprising, as the iron abundance directly affects the strength and shape of the associated spectral features. The comparatively small change in $\log(\xi)$ when the iron region is excluded suggests that information about the ionization state is distributed across a broader portion of the reflection spectrum rather than being confined to the Fe-band features. 

In the spectral plots for each source (Figure \ref{eeplots+residuals}), the \texttt{relxillCp} component contributes a larger fraction of the total model flux around the Fe band (i.e., 6-7 keV) in GRS~1716 than in GRS~1739, where the \texttt{kerrbb} contribution is comparatively stronger. We note, however, that \texttt{relxillCp} includes both the direct Comptonized continuum and the reflected emission, so this decomposition alone does not establish the relative contribution of reflection to the Fe band. This difference in the spectral shape and models offers a possible explanation for GRS 1716's preferences for high spin values, given the importance of the iron region. In contrast, the larger spin range observed in GRS 1739 reflects the importance of \texttt{kerrbb}, and hence the continuum, in modeling the iron region. This difference also explains GRS 1739's overall dependence on \texttt{kerrbb} instead of \texttt{relxillCp}, even in the high-spin fit. 

An interesting result of this experiment is the change in the reduced $\chi^2$ when the Compton-hump region is excluded. The increase in reduced $\chi^2$ indicates that the retained portion of the spectrum is described less uniformly by the restricted-band fit than the full spectrum. We do not interpret this change alone as evidence that the Compton-hump region carries greater statistical weight than the other spectral regions. Removing the Compton-hump region produces substantial changes in the constraints on $A_{\rm Fe}$ and $\log(\xi)$, both parameters of the \texttt{relxillCp} component. A smaller change is also seen in the spin constraint for the best-fit GRS~1716 solution, illustrating how broadband coverage helps restrict the range of reflection-model solutions.

The low-energy region from 0.5-3 keV, assessed by removing Swift and solely analyzing NuSTAR, has an overall minor impact on the four chosen parameters. This region of the spectrum is largely modeled by the \texttt{kerrbb} component, so its exclusion may have a more noticeable impact on the constraints of \texttt{kerrbb} parameters. Although removal of the soft band produces comparatively modest changes in several of the individual parameters examined here, its effect on the preferred solution can be substantial. For GRS~1739, the NuSTAR-only fit converges toward the alternative low-spin solution. This illustrates how a spectral region can be important for selecting between degenerate solutions even when its removal does not dramatically broaden every individual parameter constraint.

\subsubsection{Model shape vs. data}\label{subsubsec:model_shape}

\begin{figure*}[ht]
    \centering
    \includegraphics[width=0.8\textwidth]{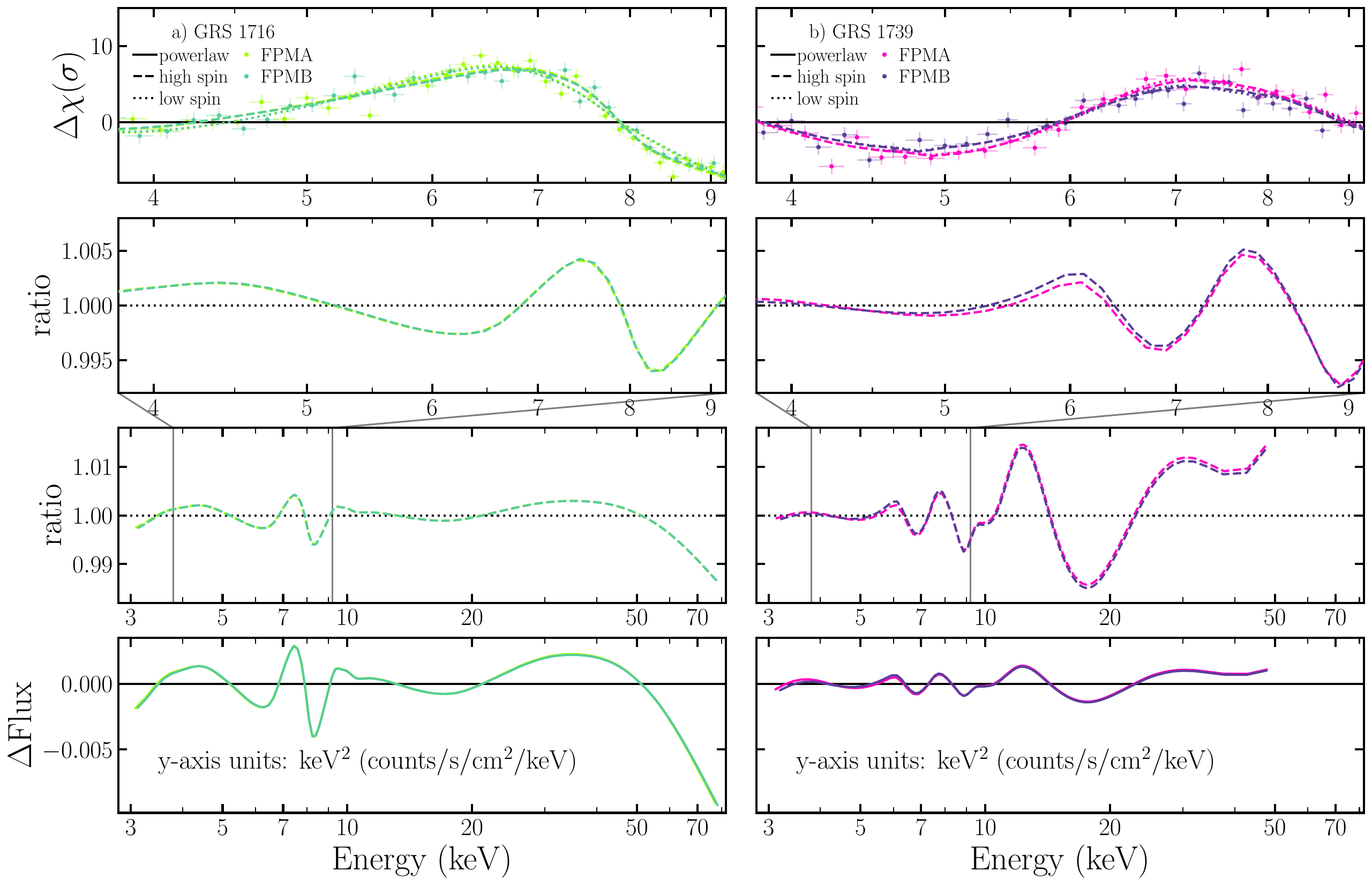}
    \caption{Top: GRS 1716 (left) and GRS 1739 (right) linked high- (dashed line) and low-spin (dotted line) models in the iron band of NuSTAR data, with respect to the residuals in terms of $\sigma$ produced by fitting the spectra with a power law (solid line). Middle: ratio of the high-spin and low-spin models in the Fe band (second row) and in the full NuSTAR band (third row). Note that for GRS 1716, the low-spin solution is $\Delta \chi^2=50$ worse than the high-spin solution, while for GRS 1739, the two solutions produce the same fit statistic. Bottom: difference in flux between the high-spin and low-spin models, in units of $\rm keV^2\;(counts/s/cm^{2}/keV)$. The ratio panels are ordered to emphasize the progression from the Fe-band region, where the high- and low-spin solutions are nearly indistinguishable, to the full NuSTAR band, where larger differences emerge at higher energies. The bottom panels then show the corresponding absolute difference in model flux, illustrating how these fractional differences translate into different statistical penalties for the two sources.}
    \label{fig:iron_line_models}
\end{figure*}

In Figure \ref{fig:iron_line_models}, we examine how the linked high- and low-spin fits for both GRS 1716 and GRS 1739 fit the iron region, given its importance in spin determination. The top panels present how the two linked spin fits and the \texttt{powerlaw} model fit the NuSTAR FPMA and FPMB spectra, and the second and third panels present the ratio of the high-spin to low-spin models for both FPMA and FPMB in the Fe band and the entire NuSTAR band, respectively. Surprisingly, despite differences in spin and $\Delta{\chi}^2=50$ in GRS 1716, both models visually fit the spectra in extremely similar manners, with differences between the two models staying within 1\% of each other in the Fe band, and barely exceeding this value above $\sim70\;\rm keV$. This discrepancy demonstrates the importance of fully exploring the extent of the parameter spaces of the models and the potential pitfall of attempting to identify solutions without proper statistical treatments. Lastly, the bottom panel in Figure \ref{fig:iron_line_models} shows the difference in flux between the high-spin and low-spin models, in units of $\rm keV^2\times(counts/s/cm^{2}/keV)$. While the ratio of the two models is smaller for GRS 1716 compared to GRS 1739, the difference in flux is significantly larger both in the Fe band and the Compton hump. Given that the source flux of GRS 1716 was higher than that of GRS 1739, for similarly small ratios of the models, the difference in flux between the models produces vastly different fit statistics for the two sources ($\Delta\chi^2=50 $ for GRS 1716 and $\Delta\chi^2=0$ for GRS 1739). This comparison provides a direct example of how data quality determines whether model flexibility is statistically penalized. Small fractional differences between competing solutions are detectable in the higher-flux GRS~1716 spectrum, whereas comparable differences in GRS~1739 can remain statistically indistinguishable, allowing the model to explore a much broader range of spin.
\\
\section{Discussion and Conclusions}
\label{sec:discussion}

We jointly modeled intermediate-state spectra of GRS 1716 and GRS 1739 using the relativistic disk model \texttt{kerrbb} and the relativistic reflection model \texttt{relxillCp}. Our primary goal was to investigate how the two spectral components interact and how correlations among their parameters affect the inferred BH spin. Within the adopted model, the GRS 1716 spectra strongly favor a high-spin solution, $a\geq0.99$, while alternative solutions produce substantially worse fit statistics. In contrast, the GRS 1739 spectra allow linked high- and low-spin solutions with nearly identical fit statistics ($a=0.97\pm0.01$ and $a=0.05^{+0.08}_{-0.24}$). Furthermore, when the spin parameters are allowed to vary separately for GRS~1739, the spin parameter associated with \texttt{relxillCp} converges toward high values, while that associated with \texttt{kerrbb} converges toward low values within the joint fit. These contrasting results show that, for these datasets and within the adopted model, the recovered spin solution is sensitive to the relative contributions of the spectral components and to how strongly the available data constrain their correlated parameters.

The different behavior of the two sources can partly be understood through the relative contributions of the \texttt{kerrbb} and \texttt{relxillCp} components across the spin-sensitive Fe-band region. In GRS~1716, \texttt{relxillCp} contributes a larger fraction of the model flux across this band, whereas in GRS~1739 the \texttt{kerrbb} contribution is comparatively stronger. Because \texttt{relxillCp} includes both the direct Comptonized continuum and the reflected emission, this decomposition should not be interpreted as a direct measure of the reflected fraction in the Fe band. The contrast between the two sources is also shaped by the constraining power of the data: as shown in Figure~\ref{eeplots+residuals} and the model-comparison experiment in Section \ref{subsubsec:model_shape}, small fractional differences between competing solutions produce a much larger absolute flux difference, and therefore a larger statistical penalty, in the brighter GRS~1716 spectrum than in GRS~1739. This combination of component contributions and data quality helps explain why departures from the preferred high-spin solution are strongly penalized for GRS~1716, while substantially different spin solutions can remain statistically comparable for GRS~1739. The high values of the \texttt{relxillCp}-associated spin parameter in both unlinked-spin fits, together with the wider range of values accessible to the \texttt{kerrbb}-associated spin parameter, further suggest that the low-spin solutions are enabled by the flexibility of the continuum component within the joint model. This conclusion applies specifically to these observations and to the adopted model configuration, and should not be interpreted as a general tendency for reflection fitting to produce higher spins than continuum fitting.

The alternative solutions arise from multidimensional parameter covariance rather than from any single two-parameter degeneracy. In \texttt{kerrbb}, changes in the BH mass, distance, mass accretion rate, spectral hardening factor, and inclination can compensate for changes in spin. As illustrated by the parameter-stepping experiment for GRS 1739 (Section \ref{subsubsec:stepping}), changing the spin alone produces a spectrum that is strongly inconsistent with the data, but adjusting the remaining continuum parameters progressively recovers a statistically equivalent solution. The reflection component exhibits a different set of correlations, involving the emissivity profile, photon index, ionization, iron abundance, disk density, and reflection fraction. These correlations allow substantially different physical descriptions to generate very similar total spectra. Indeed, the high- and low-spin models differ by less than approximately one percent across much of the Fe band, despite implying markedly different values of the spin and other model parameters (Section \ref{subsubsec:model_shape}). Statistical quality must therefore be considered together with the physical plausibility and degree of constraint of the complete parameter set. For example, in the low-spin GRS 1739 solution, the breaking radius approaches the inner disk radius, leaving only a narrow radial interval described by the inner emissivity index and driving $q_1$ toward an extreme, poorly constrained value. This behavior suggests tension within the adopted emissivity parameterization rather than providing strong evidence for a low-spin BH.

Our spectral-region experiments (Section \ref{subsec:regions}) further show that different portions of the X-ray spectrum play complementary roles in constraining the models. Removing the Fe-band region produces the largest loss of direct spin sensitivity and, in some cases, allows the fit to move toward a different spin solution. The Fe band therefore provides the most direct sensitivity to relativistic broadening and the inferred inner disk radius. The Compton hump contributes more indirectly by constraining parameters such as $\Gamma$, $\log(\xi)$, $A_{\rm Fe}$, and the reflection fraction, thereby breaking degeneracies within the reflection model and improving the stability and precision of the spin measurement. The soft X-ray band is required to characterize the thermal continuum, absorption, and the parameters entering the continuum-fitting calculation. Although removing the Swift data produced comparatively small changes in the subset of parameters examined in our band-removal experiment, fitting the GRS 1739 NuSTAR spectrum alone allowed the model to converge toward the alternative low-spin solution. Moreover, models obtained from restricted energy ranges generally failed when extrapolated across the complete spectrum. These results emphasize the importance of simultaneous broadband observations that characterize the thermal disk, Fe-band structure, and Compton hump together. Proposed missions such as HEX-P (\citealt{2024FrASS..1157834M}) or HEROIX (\citealt{heroix_paper}) demonstrate the feasibility of extending broad-band studies of BHs into the next decades. 

Independent measurements of the binary parameters and observations obtained across multiple spectral states provide complementary means of reducing these uncertainties. The continuum solution depends directly on the BH mass, distance, and inclination, and uncertainties in these quantities propagate into the inferred spin. The range of spin values obtained for GRS~1716 under different assumed distances provides a clear example of this sensitivity, with \cite{2019ApJ...887..184T} measuring a spin of $a\geq0.92$ when assuming a distance of 2.4 kpc, while \cite{2024A&A...691A.192Z} measured $a=0.464^{+0.016}_{-0.007}$ when assuming a distance of 6.9 kpc. Such external constraints should be included with their measured uncertainties rather than fixed more tightly than warranted by the data. Observations across an outburst can also isolate different parts of the spectral model: disk-dominated observations primarily constrain the thermal continuum, while observations with a stronger nonthermal component provide stronger constraints on the reflection spectrum. However, in intermediate and harder spectral states, emission associated with a compact jet may also contribute to the broadband continuum, introducing an additional source of uncertainty in separating the disk, coronal, and reflection components. We do not include an explicit jet component in the present fits, and the X-ray spectra alone do not allow us to isolate or quantify such a contribution. Any jet emission contributing over the fitted band would therefore represent an additional systematic in the decomposition of the nonthermal continuum. For GRS 1739, \cite{2024ApJ...969...40D} combined the intermediate-state spectrum studied here with a hard-state observation, and obtained an improved constraints on parameters such as the inclination, and a spin of $a=0.97^{+0.02}_{-0.07}$. More generally, simultaneous analysis of multiple spectral states can reduce the dependence of the result on the particular component balance present during a single observation (\citealt{2011MNRAS.410.2497R}).

Several limitations of the adopted spectral model should be considered when interpreting these results. First, we used \texttt{kerrbb} with the spectral hardening factor allowed to vary freely. This parameterization is useful for exposing the covariance between $f_{\rm col}$ and the inferred spin, but it does not use a disk-atmosphere calculation to restrict the hardening factor. Repeating the analysis with atmosphere-based treatments such as \texttt{kerrbb2} or \texttt{bhspec} would test whether such calculations reduce or shift the range of permitted continuum solutions. Second, our additive model, \texttt{tbabs}$\times$(\texttt{kerrbb}+\texttt{relxillCp}), treats the observed disk and coronal spectra as independent emission components. Physically, the Comptonized photons are expected to originate primarily as disk photons that are removed from the thermal spectrum and scattered by the corona. When the scattered fraction is large, neglecting this photon transfer can bias the relationship between the intrinsic disk luminosity, temperature, and inner radius on which continuum fitting relies. Convolution treatments such as \texttt{simplcut} or \texttt{thcomp} can couple the thermal and Comptonized emission more self-consistently.

A related limitation is that the reflected emission is not subsequently Compton scattered by the corona in our adopted model. If the corona intercepts a substantial fraction of the disk photons and illuminates the disk, at least some of the reflected photons may also pass through the coronal medium. Such scattering can alter the inferred reflection strength and broaden the reflected spectrum, potentially affecting constraints on the inclination, emissivity profile, and inner disk radius. The magnitude of this effect depends on the unknown coronal geometry, but it is likely to become increasingly important as the scattered fraction grows. In addition, the low-energy continuation of the Comptonized continuum and reflection spectrum is not tied self-consistently to the turnover of the disk seed-photon distribution. This can allow the nonthermal components to contribute excessive flux below or near the thermal peak, particularly for steep continua, and may introduce additional covariance with $N_{\rm H}$ and the disk parameters. Future analysis should test models that conserve the disk seed-photon budget, apply coronal scattering consistently to the direct and reflected components, and impose a low-energy turnover determined by the seed spectrum.

The scope and quality of the data in the present study introduce further limitations. We analyzed only two sources and one intermediate-state epoch for each source, and the Swift/XRT and NuSTAR observations were not simultaneous. The GRS 1716 observations were separated by eight days and show modest evolution in the soft-band flux, whereas the GRS 1739 observations were separated by three days. The Swift/XRT spectrum of GRS 1716 required the mitigation of mild pile-up, the GRS 1739 NuSTAR spectrum was affected by stray light and became background-dominated above $\sim50$ keV. These effects are particularly relevant because small changes in the recovered spectral shape can alter the relative contributions of the model components. Residual pile-up in particular could bias the soft-band continuum used to constrain \texttt{kerrbb}, while stray-light contamination and the limited high-energy SNR of GRS~1739 reduce the ability to distinguish between closely spaced spectral solutions. Studies such as the one performed in this work can benefit from including soft X-ray coverage from observatories with higher effective areas and energy resolutions (e.g., NICER, XMM-Newton, or Chandra). The mass and distance estimates also remain uncertain, particularly for GRS 1739. Finally, the analysis assumes that the disk extends to the ISCO, links the disk inclination between the continuum and reflection components, and represents the coronal illumination with prescribed emissivity profiles. Returning radiation may contribute to the illumination of the disk in intermediate and soft states, further complicating the interpretation of a phenomenological broken power-law emissivity profile \citep{2020ApJ...892...47C, 2021ApJ...909..146C, 2022MNRAS.514.3965D}. The alternative solutions identified here should therefore be regarded as demonstrations of model dependence rather than as definitive spin measurements or a complete systematic uncertainty budget.

The principal implication of this work is that, for these datasets and within the adopted modeling framework, the ability to exclude alternative spin solutions depends jointly on which spectral regions most strongly constrain the different model components and on the constraining power of the data. In intermediate-state spectra such as those studied here, the thermal, Comptonized, and reflected emission overlap over spin-sensitive energies, allowing correlated changes across a high-dimensional parameter space. For GRS~1716, the available data strongly penalize departures from the preferred high-spin solution within the adopted model. For GRS~1739, the greater flexibility permitted by the data allows substantially different spin solutions to remain statistically indistinguishable. Robust joint continuum-fitting and reflection studies therefore benefit from simultaneous broadband coverage, independent constraints on the binary properties, observations spanning multiple spectral states, physically coupled models of the disk, corona, and reflection spectrum, and thorough exploration of the full parameter space.

\begin{acknowledgments}
AZ and AO are supported by NASA under award number 80GSFC24M0006. The authors thank Ole K\"{o}nig, James Steiner, and Jon Miller for their suggestions during the early stages of the development of this manuscript. 

The authors used ChatGPT (OpenAI) during the preparation of this manuscript to assist with language editing, phrasing, and the presentation of scientific arguments. The tool was not used to perform the spectral analysis, numerical calculations, or data visualization presented in this work. All scientific content, references, interpretations, and conclusions were independently reviewed and verified by the authors, who take full responsibility for the final manuscript.
\end{acknowledgments}

\paragraph{Software}
 \texttt{Matplotlib} \citep{matplotlib}, \texttt{NumPy} \citep{numpy-guide,numpy}, \texttt{Astropy} \citep{Astropy13,astropy} 





\bibliography{paper}
\bibliographystyle{aasjournal}

\appendix

\section{Parameter Values for the GRS 1739 Stepping Video}\label{sec:par_values} 
Table \ref{stepping-endvalues} lists the two sets of parameter values, at the start and end of the stepping sequence, used to generate the video transitioning from the high-spin solution to the low-spin solution for GRS 1739. All other parameters not listed in the table were held fixed during the stepping procedure. The normalization of the \texttt{kerrbb} component was fixed to 1 for the NuSTAR FPMA spectrum, and allowed to vary for the other two spectra. The normalization of the \texttt{relxillCp} component was allowed to vary independently between the three spectra. The uncertainties reported for the MCMC columns represent the 68\% credible intervals derived from the posterior probability distributions of the parameters following the MCMC analysis. The uncertainties in the \texttt{Xspec} columns come from the square root of the diagonal entry in the covariance matrix computed by the \texttt{Xspec} fit. Many of the uncertainties are unphysical (e.g., $D_{\rm BH}=7\pm20$ kpc or $M_{\rm BH}=9.5\pm36\;\rm M_\odot$), despite the parameter space for $M_{\rm BH}$ and $D_{\rm BH}$ being restricted by hard bounds corresponding to the uncertainties adopted from the literature (e.g., the distance was constrained to $7.25\pm1.25$ kpc). This highlights that the uncertainties originating directly from \texttt{Xspec} are not sufficiently informative, and that a careful study of the parameter space is required. We note that the exercise of parameter stepping does not take into account the parameter uncertainties, just the values.

\begin{table}[htbp]
\centering
\small
\caption{Parameter values for the high-spin and low-spin fits of GRS 1739
\label{stepping-endvalues}}
\resizebox{\columnwidth}{!}{%
\renewcommand{\arraystretch}{1.4}
\begin{tabular}{l|cc|cc}
\hline
Parameter &
High Spin \texttt{Xspec} &
High Spin MCMC &
Low Spin \texttt{Xspec} &
Low Spin MCMC \\
\hline
$N_{\rm H}\;(\times10^{22}\;\rm cm^{-2})$ & $2.9\pm0.1$ & $2.894^{+0.005}_{-0.128}$ & $2.85\pm0.09$ & $2.83^{+0.02}_{-0.06}$ \\
$M_{\rm BH}\;(M_\odot)$ & $9.5\pm36$ & $9.5^\dagger$ & $5\pm24$ & $7.8^{+0.8}_{-1.4}$ \\
$\dot{M}\;(\times10^{18}\;\rm g\;s^{-1})$ & $0.1\pm0.8$ & $0.13^{+0.03}_{-0.01}$ & $1\pm4$ & $0.64^{+0.17}_{-0.07}$ \\
$D_{\rm BH}\;(\rm kpc)$ & $7\pm20$ & $6.9^{+0.6}_{-0.4}$ & $8\pm20$ & $6.27^{+1.56}_{-0.03}$ \\
$f$ & $1.4\pm1.2$ & $1.40^{+0.05}_{-0.09}$ & $1.7\pm4.6$ & $2.3\pm0.2$ \\
$\theta\;(^\circ)$ & $63\pm8$ & $62^{+3}_{-2}$ & $72\pm33$ & $76^{+1}_{-6}$ \\
$a$ & $0.97\pm0.04$ & $0.974^{+0.004}_{-0.008}$ & $0.1\pm0.6$ & $0.05^{+0.08}_{-0.24}$ \\
$R_{\rm br}\;(r_{\rm g})$ & $3.5\pm2.8$ & $3.8\pm0.5$ & $6\pm120$ & $6^{+1}_{-3}$ \\
$q_{\rm 1}$ & $9.7\pm6.7$ & $8.2^{+1.5}_{-0.4}$ & $10\pm7500$ & $9.7^{+0.3}_{-2.6}$ \\
$q_{\rm 2}$ & $2.1\pm0.6$ & $2.1^{+0.2}_{-0.3}$ & $2.2\pm0.4$ & $2.04^{+0.20}_{-0.02}$ \\
$\Gamma$ & $2.80\pm0.01$ & $2.804^{+0.002}_{-0.001}$ & $2.1\pm0.2$ & $2.1^{+0.02}_{-0.03}$ \\
$\log(\xi)\;(\rm erg\;cm\;s^{-1})$ & $3.4\pm0.2$ & $3.39^{+0.01}_{-0.02}$ & $4.1\pm0.7$ & $4.03^{+0.18}_{-0.03}$ \\
$\log(n)\;(\rm cm^{-3})$ & $19.1\pm0.3$ & $19.07\pm0.01$ & $15\pm28$ & $15.2^{+0.9}_{-0.2}$ \\
$A_{\rm Fe}\;(A_\odot)$ & $6\pm7$ & $5.6^{+0.4}_{-0.3}$ & $6\pm13$ & $5.0^{+2.0}_{-0.7}$ \\
$kT_{\rm e}\;(\rm keV)$ & $300\pm4000$ & $385^{+15}_{-146}$ & $79\pm145$ & $113^{+40}_{-64}$ \\
$\mathrm{refl\_frac}$ & $3\pm14$ & $3.4^{+0.7}_{-0.4}$ & $3\pm4$ & $2.0^{+1.1}_{-0.2}$ \\
$\mathrm{norm}_{\rm k,FPMA}$ & $1^*$ & $1^*$ & $1^*$ & $1^*$ \\
$\mathrm{norm}_{\rm k,FPMB}$ & $0.930\pm0.007$ & $0.931^{+0.005}_{-0.001}$ & $0.932\pm0.004$ & $0.933\pm0.004$ \\
$\mathrm{norm}_{\rm k,XRT}$ & $1.15\pm0.03$ & $1.15^{+0.01}_{-0.04}$ & $1.31\pm0.03$ & $1.14\pm0.02$ \\
$\mathrm{norm}_{\rm r,FPMA}\;(\times10^{-4})$ & $3\pm10$ & $2.5^{+0.7}_{-0.2}$ & $8\pm9$ & $7.9^{+2.9}_{-0.4}$ \\
$\mathrm{norm}_{\rm r,FPMB}\;(\times10^{-4})$ & $3\pm10$ & $2.5^{+0.7}_{-0.2}$ & $8\pm8$ & $7.7^{+3.1}_{-0.2}$ \\
$\mathrm{norm}_{\rm r,XRT}\;(\times10^{-4})$ & $3\pm10$ & $3.25^{+0.01}_{-0.08}$ & $9\pm9$ & $8.6^{+2.7}_{-1.4}$ \\
\hline
$\chi^2\; (\nu=392)$ & 497 & $510^{+6}_{-5}$ & 497 & $510^{+6}_{-4}$ \\
\hline
\end{tabular}
}
\begin{minipage}{\columnwidth}
\scriptsize
$^*$ indicates a fixed parameter. The normalization of the \texttt{kerrbb} component for FPMA was fixed to 1. $^\dagger$ indicates a free parameter that was pegged at one of the edges of the parameter space in the MCMC run.
\end{minipage}
\end{table}

\newpage
\section{Corner Plots}\label{sec:corner_plots}

\begin{figure*}[ht!]
\centering
\includegraphics[width=1.0\linewidth]{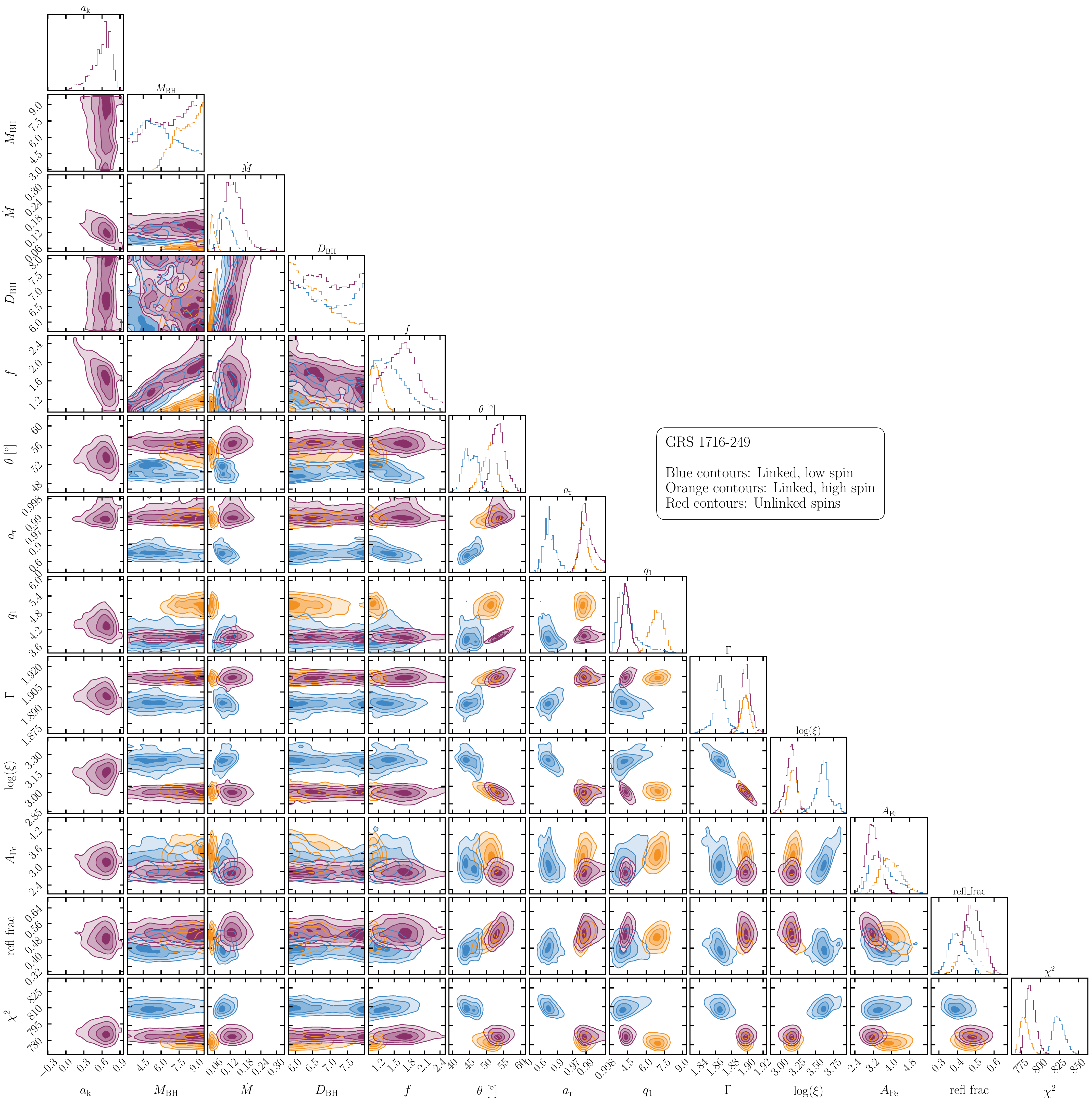}
\caption{Corner plot for GRS 1716 highlighting the behavior of a few parameters of interest in the \texttt{kerrbb} and \texttt{relxillCp} components. The red contours show the solutions with the \texttt{kerrbb} and \texttt{relxillCp} spins unlinked. The orange and blue contours show the solutions with linked high and low spins, respectively. The selection of these parameters was made based on their correlations with the spin, as described in Appendix \ref{sec:regression_results}. The subscripts $k$ and $r$ on the spin values reported (first and seventh columns) indicate the model that the spin is measured by, representing the \kerrbb and \relxillcp models. When the spins are linked between the models, we only show the contours in the $a_r$ column. The inclination was linked between the \kerrbb and \relxillcp components. While for GRS 1716 we allowed the power-law index $\Gamma$ to vary between \nustar and Swift/XRT, this figure shows the values measured by the \nustar spectra. }
\label{fig:corner_1716}
\end{figure*}

\begin{figure*}[ht!]
\centering  
\includegraphics[width=1.0\linewidth]{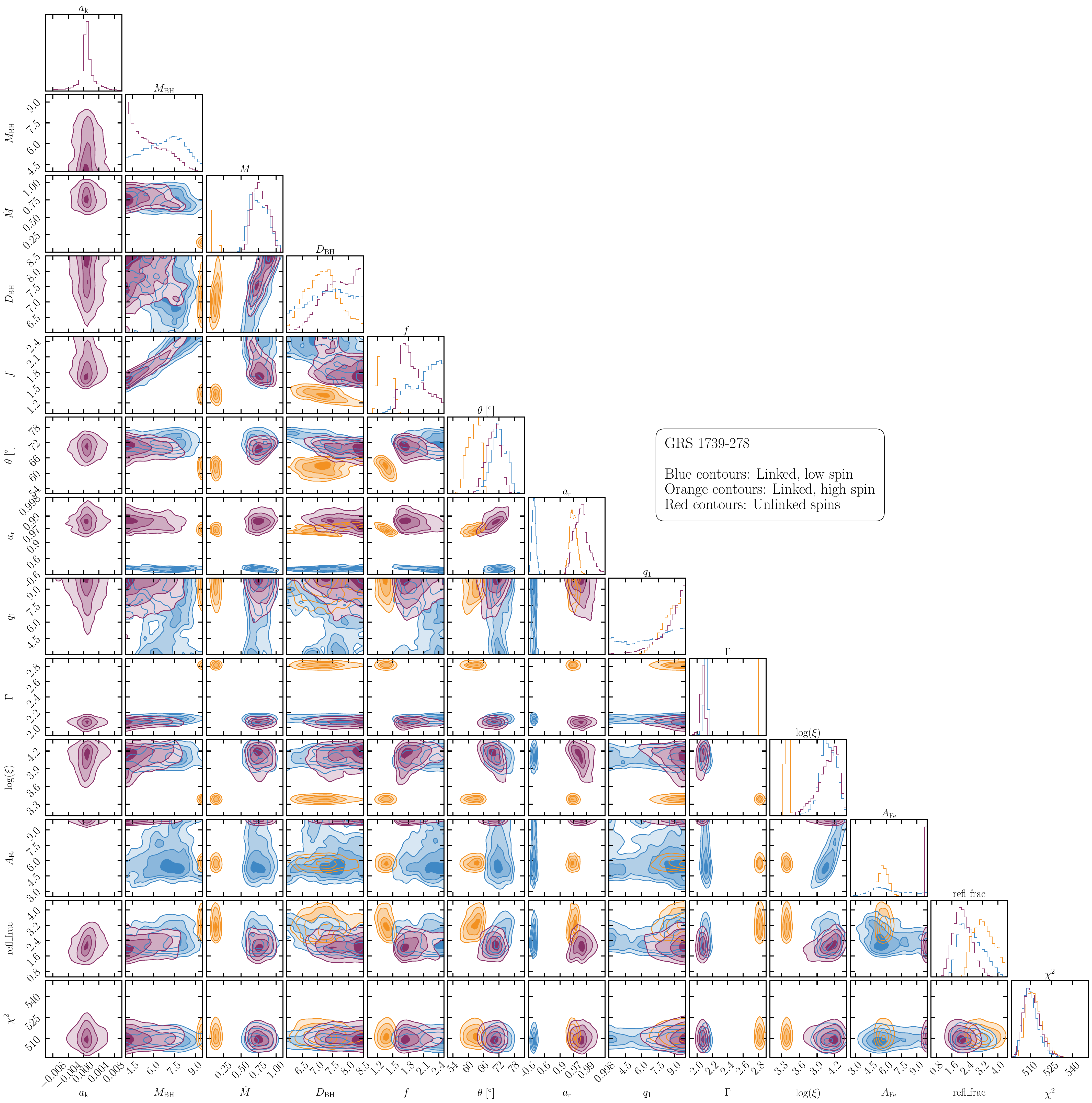}
\caption{Corner plot for GRS 1739 highlighting the behavior of a few parameters of interest in the \texttt{kerrbb} and \texttt{relxillCp} components. The red contours show the solutions with the \texttt{kerrbb} and \texttt{relxillCp} spins unlinked. The orange and blue contours show the solutions with linked high and low spins, respectively. The selection of these parameters was made based on their correlations with the spin, as described in Appendix \ref{sec:regression_results}. The subscripts $k$ and $r$ on the spin values reported (first and seventh columns) indicate the model that the spin is measured by, representing the \kerrbb and \relxillcp models. When the spins are linked between the models, we only show the contours in the $a_r$ column. The inclination was linked between the \kerrbb and \relxillcp components.}
\label{fig:corner_1739}
\end{figure*}

Figures \ref{fig:corner_1716} and \ref{fig:corner_1739} show the corner plots obtained based on the three fits to the GRS 1716 and GRS 1739 spectra, respectively. The red contours show the posterior distributions for the fits in which the spins were unlinked between the \texttt{kerrbb} and \texttt{relxillCp} components, and the orange and blue contours show the posteriors for the high-spin and low-spin solutions, respectively, when the spin was linked between the two model components. 

\newpage
\section{Multi-parameter regression}\label{sec:regression_results}

Based on the MCMC analysis, we construct contour plots to visualize the relationship between parameter pairs, with contour lines containing equal probabilities of parameter combinations. While it is possible to identify two-parameter degeneracies visually from contour plots, there are useful mathematical measures for quantitatively determining the degree to which two parameters are correlated. One widely used measure is the Pearson correlation coefficient $\rho$, which quantifies the strength and direction of a linear relationship between two variables, $x$ and $y$. More specifically, $\rho$ is defined as the covariance of $x$ and $y$  divided by the product of their standard deviations. The coefficient captures the degree to which deviations from the mean in one variable correspond to deviations in the other. 
\begin{equation}
\rho(x,y) = \frac{\sum_{i=1}^{n}(x_i - \bar{x})(y_i - \bar{y})}{\sigma_x\sigma_y}  
\end{equation}
The closer the absolute value of the Pearson coefficient is to 1, the stronger the linear relationship between the two parameters. The sign of $\rho$ indicates the direction of the correlation: a positive value corresponds to a generally upward-sloping contour plot, where as one parameter increases, the other must also increase to compensate for changes in the spectral shape. A negative value indicates an inverse relationship between the parameters, where one must increase and the other decrease to create a spectral fit of similar quality.

Although the Pearson correlation coefficient is useful for quantifying correlation between pairs of parameters, it is inherently limited to two-dimensional analysis. However, as highlighted in Figure \ref{high_to_low}, degeneracies can also arise among larger sets of parameters. Another tool for understanding these higher-dimensional correlations is weighted least-squares regression, which can be used to evaluate the dependence of one variable on a combination of multiple independent variables. Weighted least-squares regression finds the best-fit linear model to two or more sets of data by minimizing a weighted sum of the squared errors. For one dependent variable $z$ and two independent variables $x$ and $y$, the weighted squared error (WSE) is given by the following expression: 
\begin{equation}
\text{WSE} = \sum_{i=1}^{n} w_i \left( (m_x x_i + m_y y_i + b) - z_i \right)^2  
\end{equation}
The weighting $w_i$ varies for every data point and can be used to emphasize some spectral fits over others to determine the best-fit line. We apply $1/{{\chi}^2}$ weighting in our analysis to assign more importance to higher probability fits associated with lower ${\chi}^2$. Taking the first derivative of the WSE, it is possible to determine the slope values and intercepts that define the best-fit line. The quality of this regression fit is assessed using the coefficient of determination, $R^2$, which quantifies how well the variation in the dependent variable can be explained by variation in the independent variables. $R^2$ is calculated using the ratio of the WSE to the total squared error (TSE), which is defined as the sum of the squared deviations of $z$ from its mean:
\begin{equation}
TSE = \sum_{i=1}^{n}{(z_i-\bar{z})^2}
\end{equation}
$R^2$ is then computed as:
\begin{equation}
R^2 = 1 - \frac{\text{WSE}}{TSE}
\end{equation}
Values of $R^2$ closest to 1 indicate that a large proportion of the variation in the dependent variable can be explained by the other independent variables, implying a high amount of degeneracy between the combinations of parameters. Conversely, values closer to 0 suggest greater independence of the parameters. Unlike the Pearson $\rho$, values of $R^2$ are only positive and therefore do not indicate the direction of correlation, only the strength. To perform these calculations on our MCMC runs and compute $R^2$ for various parameter combinations, we used functions from the Python library \texttt{statsmodels} (\citealt{seabold2010statsmodels}).

\begin{figure*}
\centering
\begin{minipage}[t]{0.58\textwidth}
    \centering
    \vspace{0pt}
    \includegraphics[width=\linewidth]{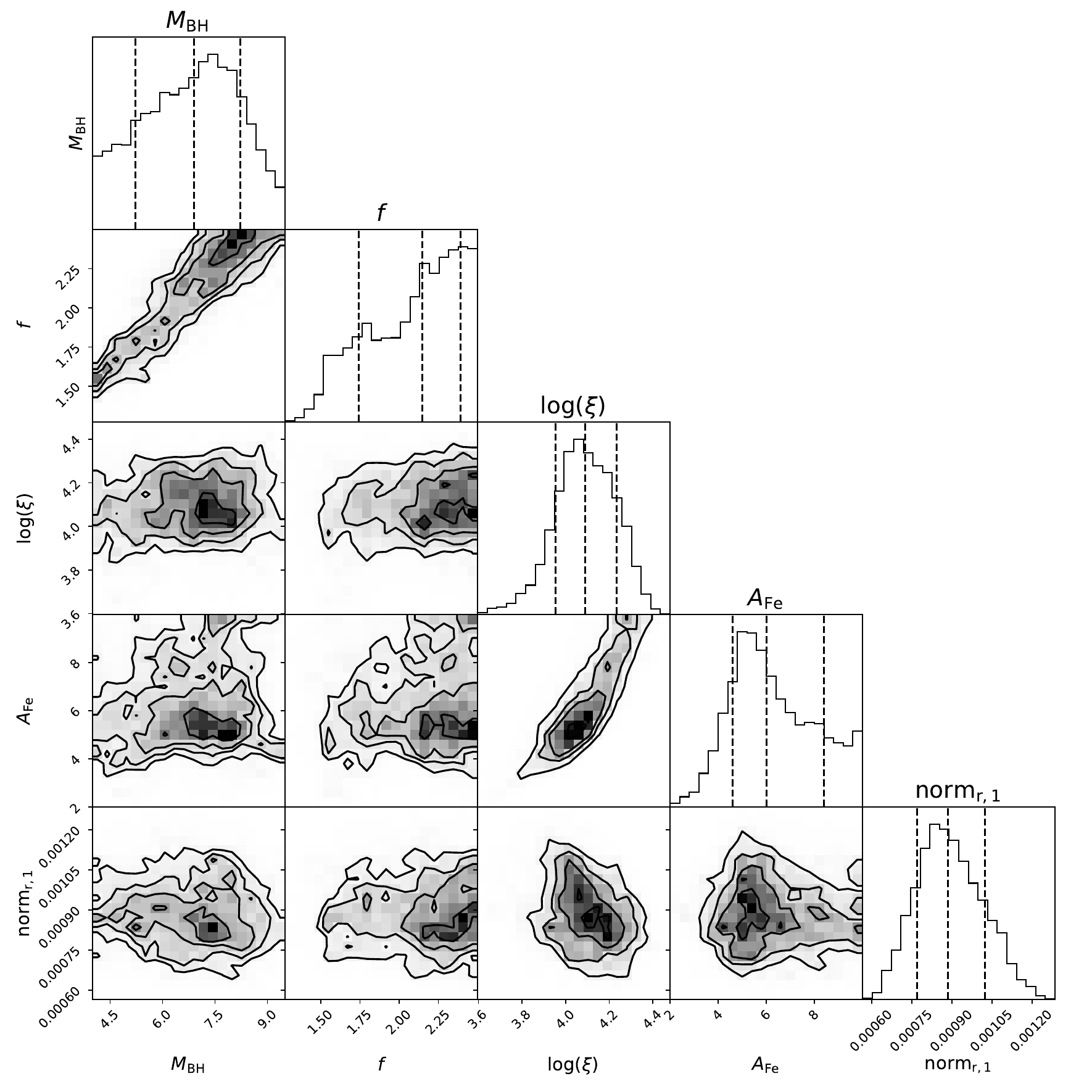}
\end{minipage}%
\hfill
\begin{minipage}[t]{0.38\textwidth}
    \vspace{0pt}
    \centering
    {\renewcommand{\arraystretch}{2.8}\footnotesize
    \resizebox{\linewidth}{!}{%
    \begin{tabular}{| l | l | c | c |}
    \hline
    Parameter 1 & Parameter 2 & $\rho$ & $R^2$ \\
    \hline
    $M_{\rm BH}$ & $f$ & 0.872 & 0.761 \\ \hline
    $\log(\xi)$ & $A_{\rm Fe}$ & 0.816 & 0.666 \\ \hline
    $\log(\xi)$ & $\mathrm{norm}_{r,1}$ & $-0.303$ & 0.091 \\ \hline
    $M_{\rm BH}$ & $A_{\rm Fe}$ & 0.159 & 0.025 \\ \hline
    $M_{\rm BH}$ & $\log(\xi)$ & 0.158 & 0.025 \\ \hline
    $f$ & $\log(\xi)$ & 0.148 & 0.022 \\ \hline
    $f$ & $A_{\rm Fe}$ & 0.136 & 0.018 \\ \hline
    $M_{\rm BH}$ & $\mathrm{norm}_{r,1}$ & $-0.048$ & 0.002 \\ \hline
    $A_{\rm Fe}$ & $\mathrm{norm}_{r,1}$ & $-0.025$ & 0.001 \\ \hline
    $f$ & $\mathrm{norm}_{r,1}$ & $-0.000$ & 0.000 \\ \hline
    \end{tabular}%
    }}
\end{minipage}

\caption{A condensed corner plot for the GRS~1739 low-spin solution, highlighting five parameters and their associated Pearson correlation coefficients ($\rho$) and coefficients of determination ($R^2$). Strong correlations are reflected in the linearity of the contours (e.g., $M_{\rm BH}-f$ and $log(\xi)-A_{\rm Fe}$), while others are weak or absent.}
\label{top_5_low_spin}
\end{figure*}

While $\rho$ and $R^2$ may yield different values for the same combination of parameters, both metrics effectively measure the same underlying concept of degeneracy, and they produce essentially identical rankings given the strength of parameter pairings. As an example, Figure \ref{top_5_low_spin} highlights 5 degenerate parameters and their corresponding coefficient values. The most correlated parameter pairs, which appear as distinctly elongated linear contours, are the BH mass $M_{\rm BH}$ with the hardening factor $f$, and the ionization $\log(\xi)$ with the Fe abundance $A_{\rm Fe}$.

Unlike with simple two-parameter contour plots, it is not easy, or even possible, to visualize degeneracies in higher dimensions. For these complicated relationships, we rely on the coefficient of determination from weighted least-squares regression as a quantitative measure of parameter combination dependencies. In our analysis, we set the spin parameter as our dependent variable and run linear regression with one to four other variables, saving the top three combinations with the highest $R^2$ values. 

Table \ref{regression-results} details the three highest performing parameter combinations of one to four independent parameters and their associated $R^2$ values for both GRS 1716 and GRS 1739. The $R^2$ values were determined by weighted least-squares regression on the spin parameter in the three different spin solutions: linked high spin, unlinked spin (combinations for the \texttt{kerrbb} and \texttt{relxillCp} model components are in separate columns), and linked low spin.

\begin{sidewaystable*}[p]
\centering
\scriptsize
\renewcommand{\arraystretch}{1.5}
\caption{Weighted Least-Squares Regression Results}
\label{regression-results}
\resizebox{\textheight}{!}{%
\begin{tabular}{lllll}
\hline
\multirow{2}{*}{Source} & \multirow{2}{*}{Linked High Spin} & \multicolumn{2}{c}{Unlinked Spin} & \multirow{2}{*}{Linked Low Spin}\\ 
 &  & \texttt{kerrbb} & \texttt{rexillCp} & \\
\hline
\multirow{12}{*}{GRS 1716} & refl\_frac (0.264)                                        & $\dot{M}$ (0.460)                                        & refl\_frac,  (0.123)                                       & $\theta\;[^\circ]$ (0.498)                               \\
                           & $A_{\rm Fe}$ (0.233)                                      & $f$ (0.350)                                              & $A_{\rm Fe}$ (0.067)                                       & $f$ (0.462)                                              \\
                           & $f$ (0.167)                                               & $q_1$ (0.036)                                            & $N_{\rm H}$ (0.0627)                                       & $\log(\xi)$ (0.287)                                      \\
                           & $A_{\rm Fe}$, refl\_frac (0.670)                          & $\dot{M}$, $D_{\rm BH}$ (0.863)                          & $A_{\rm Fe}$, refl\_frac (0.405)                           & $\theta\;[^\circ]$, $q_1$ (0.866)                        \\
                           & $\theta\;[^\circ]$, $A_{\rm Fe}$ (0.409)                  & $\dot{M}$, $f$ (0.725)                                   & refl\_frac, $\rm norm_{r,1}$ (0.204)                       & $\dot{M}$, $D_{\rm BH}$ (0.793)                          \\
                           & refl\_frac, $\rm norm_{r,1}$ (0.389)                      & $M_{\rm BH}$, $f$ (0.694)                                & $N_{\rm H}$, refl\_frac (0.150)                            & $M_{\rm BH}$, $f$ (0.723)                                \\
                           & $q_1$, $A_{\rm Fe}$, refl\_frac (0.849)                   & $M_{\rm BH}$, $\dot{M}$, $f$ (0.964)                     & $q_1$, $A_{\rm Fe}$, refl\_frac (0.493)                    & $M_{\rm BH}$, $\dot{M}$, $f$ (0.936)                     \\
                           & $A_{\rm Fe}$, refl\_frac, $\rm norm_{k,3}$ (0.731)        & $N_{\rm H}$, $\dot{M}$, $D_{\rm BH}$ (0.915)             & $\theta\;[^\circ]$, $A_{\rm Fe}$, refl\_frac (0.469)       & $\theta\;[^\circ]$, $q_1$, $A_{\rm Fe}$ (0.901)          \\
                           & $R_{\rm br}$, $A_{\rm Fe}$, refl\_frac (0.720)            & $\dot{M}$, $D_{\rm BH}$, $\rm norm_{k,3}$, (0.910)       & $A_{\rm Fe}$, refl\_frac, $\rm norm_{k,3}$ (0.426)         & $f$, $\theta\;[^\circ]$, $q_1$ (0.881)                   \\
                           & $q_1$, $A_{\rm Fe}$, refl\_frac, $\rm norm_{r,1}$ (0.886) & $M_{\rm BH}$, $\dot{M}$, $D_{\rm BH}$, $f$ (0.985)       & $q_1$, $\log(\xi)$, refl\_frac, $\rm norm_{r,1}$ (0.804)   & $M_{\rm BH}$, $\dot{M}$, $D_{\rm BH}$, $f$ (0.966)       \\
                           & $q_1$,  $\Gamma$, $A_{\rm Fe}$, refl\_frac (0.876)        & $M_{\rm BH}$, $\dot{M}$, $f$, $q_1$ (0.974)              & $q_1$, $A_{\rm Fe}$, refl\_frac, $\rm norm_{r,1}$ (0.779)  & $M_{\rm BH}$, $\dot{M}$, $f$, $q_1$ (0.959)              \\
                           & $q_1$, $\log(\xi)$, $A_{\rm Fe}$, refl\_frac (0.869)      & $M_{\rm BH}$, $\dot{M}$, $f$, $\theta\;[^\circ]$ (0.972) & $q_1$, $\Gamma$, $A_{\rm Fe}$, refl\_frac (0.727)          & $M_{\rm BH}$,  $\dot{M}$, $f$, $\rm norm_{k,3}$ (0.953)  \\
\hline
\multirow{12}{*}{GRS 1739} & $f$ (0.580)                                               & $kT_{\rm e}$ (0.014)                                     & $\theta\;[^\circ]$ (0.352)                                 & $f$ (0.355)                                              \\
                           & $q_2$ (0.358)                                             & $q_1$ (0.010)                                            & $\log(\xi)$ (0.132)                                        & $\dot{M}$ (0.197)                                        \\
                           & $\theta\;[^\circ]$ (0.222)                                & refl\_frac (0.009)                                       & $q_1$ (0.094)                                              & $R_{\rm br}$ (0.091)                                     \\
                           & $\dot{M}$, $D_{\rm BH}$ (0.942)                           & $f$, $kT_{\rm e}$ (0.025)                                & $\theta\;[^\circ]$, $q_1$ (0.603)                          & $M_{\rm BH}$, $f$ (0.736)                                \\
                           & $\dot{M}$, $f$ (0.813)                                    & $a_{\rm r}$, $kT_{\rm e}$ (0.024)                        & $\dot{M}$, $D_{\rm BH}$ (0.385)                            & $\dot{M}$, $f$ (0.593)                                   \\
                           & $D_{\rm BH}$, $f$ (0.707)                                 & $\Gamma$,  $kT_{\rm e}$ (0.024)                          & $N_{\rm H}$, $\theta\;[^\circ]$ (0.373)                    & $\dot{M}$, $D_{\rm BH}$ (0.465)                          \\
                           & $\dot{M}$, $f$, $\theta\;[^\circ]$ (0.965)                & $M_{\rm BH}$, $\dot{M}$, $f$ (0.038)                     & $\theta\;[^\circ]$, $q_1$, $\log(\xi)$ (0.674)             & $M_{\rm BH}$, $\dot{M}$, $f$ (0.978)                     \\
                           & $N_{\rm H}$, $\dot{M}$, $D_{\rm BH}$ (0.955)              & $f$, $\Gamma$, $kT_{\rm e}$ (0.038)                      & $\dot{M}$, $D_{\rm BH}$, $q_1$ (0.653)                     & $\dot{M}$, $D_{\rm BH}$, $\theta\;[^\circ]$ (0.956)      \\
                           & $\dot{M}$, $D_{\rm BH}$, $f$ (0.951)                      & $a_{\rm r}$, $\Gamma$, $kT_{\rm e}$ (0.036)              & $\theta\;[^\circ]$, $q_1$, $q_2$ (0.623)                   & $M_{\rm BH}$, $D_{\rm BH}$, $f$ (0.804)                  \\
                           & $\dot{M}$, $D_{\rm BH}$, $f$, $\theta\;[^\circ]$ (0.984)  & $M_{\rm BH}$, $\dot{M}$, $D_{\rm BH}$ $f$ (0.090)        & $q_1$, $\Gamma$, refl\_frac, $\rm norm_{r,1}$ (0.743)      & $M_{\rm BH}$, $\dot{M}$, $D_{\rm BH}$, $f$ (0.989)       \\
                           & $N_{\rm H}$, $\dot{M}$, $f$, $\theta\;[^\circ]$ (0.984)   & $M_{\rm BH}$, $\dot{M}$, $f$, $\theta\;[^\circ]$ (0.072) & $q_1$, $\log(\xi)$, refl\_frac, $\rm norm_{r,1}$ (0.742)   & $M_{\rm BH}$, $\dot{M}$, $f$, $\theta\;[^\circ]$ (0.987) \\
                           & $N_{\rm H}$, $\dot{M}$, $f$, $\theta\;[^\circ]$ (0.971)   & $M_{\rm BH}$, $\dot{M}$, $f$, $kT_{\rm e}$ (0.064)       & $\theta\;[^\circ]$, $q_1$, $\log(\xi)$, refl\_frac (0.735) & $M_{\rm BH}$, $\dot{M}$, $f$, refl\_frac (0.982) \\
\hline
\end{tabular}
}

\begin{minipage}{\textheight}
\scriptsize
$^*$ Results of weighted least-squares regression on the spin parameter with up to 4 independent variables. The top 3 variable combinations with the highest $R^2$ are reflected for each number of independent variables.
\end{minipage}
\end{sidewaystable*}

\end{document}